\documentclass[aip,jcp,amsmath,amssymb,floatfix,reprint,citeautoscript,noeprint]{revtex4-1}

\usepackage[utf8]{inputenc}
\usepackage{graphicx}
\usepackage{wrapfig}
\usepackage[colorlinks,allcolors=black,citecolor=blue,urlcolor=blue]{hyperref}
\usepackage{chemformula}
\usepackage{placeins}

\graphicspath{{figures/}}

\begin{document}

\def\mytitle{
Defect-Dependent Corrugation in Graphene
}
\title{\mytitle}

\author{Fabian L. Thiemann}%
\affiliation{%
Department of Physics and Astronomy, University College London, Gower Street, London, WC1E 6BT, United Kingdom
}%
\affiliation{%
Thomas Young Centre and London Centre for Nanotechnology, 17-19 Gordon Street, London WC1H 0AH, United Kingdom
}%
\affiliation{%
Yusuf Hamied Department of Chemistry, University of Cambridge, Lensfield Road, Cambridge, CB2 1EW, United Kingdom
}
\affiliation{%
Department of Chemical Engineering, Imperial College London,
South Kensington Campus, London SW7 2AZ, United Kingdom
}%
\author{Patrick Rowe}%
\affiliation{%
Department of Physics and Astronomy, University College London, Gower Street, London, WC1E 6BT, United Kingdom
}%
\affiliation{%
Thomas Young Centre and London Centre for Nanotechnology, 17-19 Gordon Street, London WC1H 0AH, United Kingdom
}%
\affiliation{%
Yusuf Hamied Department of Chemistry, University of Cambridge, Lensfield Road, Cambridge, CB2 1EW, United Kingdom
}
\author{Andrea Zen}
\affiliation{Dipartimento di Fisica Ettore Pancini, Universit\`a di Napoli Federico II, Monte S. Angelo, I-80126 Napoli, Italy}
\affiliation{Department of Earth Sciences, University College London, Gower Street, London WC1E 6BT, United Kingdom}
\affiliation{%
Thomas Young Centre and London Centre for Nanotechnology, 17-19 Gordon Street, London WC1H 0AH, United Kingdom
}%
\author{Erich A. Müller}%
\affiliation{%
Department of Chemical Engineering, Imperial College London,
South Kensington Campus, London SW7 2AZ, United Kingdom
}%
\author{Angelos Michaelides}%
\email{am452@cam.ac.uk}
\affiliation{%
Yusuf Hamied Department of Chemistry, University of Cambridge, Lensfield Road, Cambridge, CB2 1EW, United Kingdom
}
\affiliation{%
 Department of Physics and Astronomy, University College London, Gower Street, London, WC1E 6BT, United Kingdom
}%
\affiliation{%
Thomas Young Centre and London Centre for Nanotechnology, 17-19 Gordon Street, London WC1H 0AH, United Kingdom
}

\begin{abstract}
Graphene's intrinsically corrugated and wrinkled topology fundamentally influences its electronic, mechanical, and chemical properties.
Experimental techniques allow the manipulation of pristine graphene and the controlled production of defects which allows to control the atomic out-of-plane fluctuations and, thus, tune graphene's properties. 
Here, we perform large scale machine learning-driven molecular dynamics simulations to understand the impact of defects on the structure of graphene.
We find that defects cause significantly higher corrugation leading to a strongly wrinkled surface.
The magnitude of this structural transformation strongly depends on the defect concentration and specific type of defect.
Analysing the atomic neighborhood of the defects reveals that the extent of these morphological changes depends on the preferred geometrical orientation and the interactions between defects.
While our work highlights that defects can strongly affect graphene's morphology, it also emphasises the differences between distinct types by linking the global structure to the local environment of the defects.
\end{abstract}

\date{This manuscript was compiled on \today}

{\maketitle}

Graphene's remarkable properties strongly depend on its morphology with many physical phenomena arising as a consequence of its intrinsic ripples \cite{Fasolino2007} and corrugation \cite{Qin2016,Guinea2008,Guinea2008a, VazquezDeParga2008,Boukhvalov2009,Qiu2011,Wei2014,Dahanayaka2017,Marbach2018} or by the presence of defects \cite{Lehtinen2003,Sammalkorpi2004,CostaFilho2007,Kopylov2011,Wang2012,Ansari2012, Mortazavi2013, He2014, Zandiatashbar2014, Wei2012,Boukhvalov2008,Banhart2011,Nair2012,Malekpour2016,Valencia2017}.
For instance, introducing eight-membered-ring defects can enhance graphene’s ion permeability \cite{Griffin2020,Hu2014a,Sun2020}.
This makes graphene an ideal candidate for nanoengineering where material properties are tuned by modifying the atomic morphology \cite{Lusk2008,Krasheninnikov2007}.
To this end, a plethora of experimental techniques has emerged ranging from the atomically precise insertion of defects via electron beam \cite{Krasheninnikov2001,Krasheninnikov2007,Kotakoski2011,Kotakoski2011a,Robertson2012,Susi2017,Zhao2017,Tripathi2018, Trentino2021} or ion bombardment\cite{Krasheninnikov2001,Yazyev2007,Yoon2016,Shi2019,Su2021} to chemical etching with KOH \cite{Zhu2011, Wang2016} and the regulation of rippling patterns by inducing strain \cite{Bao2009}.
More recent approaches like laser-assisted chemical vapour deposition \cite{Toh2020} or high temperature quenching \cite{Zhao2019} go one step further by incorporating the desired morphology \textit{a priori} in the growth process.\\

The diversity of methods available to manipulate graphene's atomic structure highlights the potential of morphologically-modified graphene which is also reflected by the variety of exciting applications.
In particular, our interest has been piqued by reports on the tuned capability of graphene in the fields of hydrogen storage \cite{Tozzini2011,Guo2012}, catalysis \cite{Ito2016}, and ultrahigh and fast adsorption of organic pollutants \cite{Wang2014,Wang2016}.
Introducing defects can significantly increase graphene's number of active sites due to both the reactive character of the defects itself \cite{Boukhvalov2008,Ito2016, Wang2016} and the induced conformational transformation to a highly corrugated surface where wrinkles possess a high chemical activity \cite{Boukhvalov2009,Glukhova2012,Chen2015}. 
The rippled and curled shape, moreover, ensures that the created active sites are accessible for adsorbates or chemical dopants \cite{Ito2016} by expanding the surface area and preventing graphene sheets from stacking together \cite{Guo2012,Wang2016}. 
It was also suggested, that the adsorption affinity of molecules is directly linked to the corrugation profile \cite{Radich2013,Chen2015,Wang2016} indicating the opportunity for the selective removal of organic pollutants.\\

The impact of defects on graphene's structure is at the very heart of the applications described above raising the question of distinct deformation mechanisms.
It is well known that even small deviations from a pristine hexagonal lattice, such as isolated pentagons or heptagons, can lead to a corrugated surface \cite{Terrones1992,Irvine2010,Liu2010,Kawasumi2013,Kusumaatmaja2013,Brojan2015}.
These induced morphological changes have been investigated for a variety of defects in graphene including dislocations (pentagon-heptagon dipoles) \cite{Warner2013,Lehtinen2013,Jain2015}, vacancies \cite{Schniepp2008,Kotakoski2014}, topological defects \cite{Samsonidze2002,Pang2019}, adatoms \cite{Lehtinen2015}, and grain boundaries \cite{Yazyev2010,Liu2011,Hofer2018}.
Several of these experimental \cite{Warner2012,Lehtinen2013} and computational \cite{Samsonidze2002,Yazyev2010,Kotakoski2014,Jain2015} studies report a long-ranging interaction between defects which manifests itself in an out-of-plane buckling of the surface.
These deformations are dictated by the type and arrangement of defects which has lead to fascinating work aiming to predict the unique arrangement of defects required to obtain a desired three-dimensional structure \cite{Wang2013,Zhang2014,Zhang2014a}.
Notwithstanding the value of these studies, fundamental questions such as whether the magnitude of the induced corrugation varies between distinct types of defects or how the morphology changes for a highly defective system have yet to be systematically addressed.
In fact, exploring the dependence of graphene's surface roughness on both the nature and number of defects present and revealing the specific mechanisms responsible for the structural transformation are of both theoretical and practical interest.
In particular, it may contribute to a better understanding of the puzzling experiments reporting an enhancement of graphene's stiffness\cite{Lopez-Polin2015} and a vanishing thermal expansion coefficient\cite{Lopez-Polin2017} with an increasing vacancy concentration.
Knowing the impact of defects on graphene's morphology \textit{a priori}, would be a powerful tool to further tailor graphene's properties and, thus, accelerate the design and manufacture of graphene-based nanomaterials.\\
\begin{figure}[t!]
\centering
\includegraphics[width=1\linewidth]{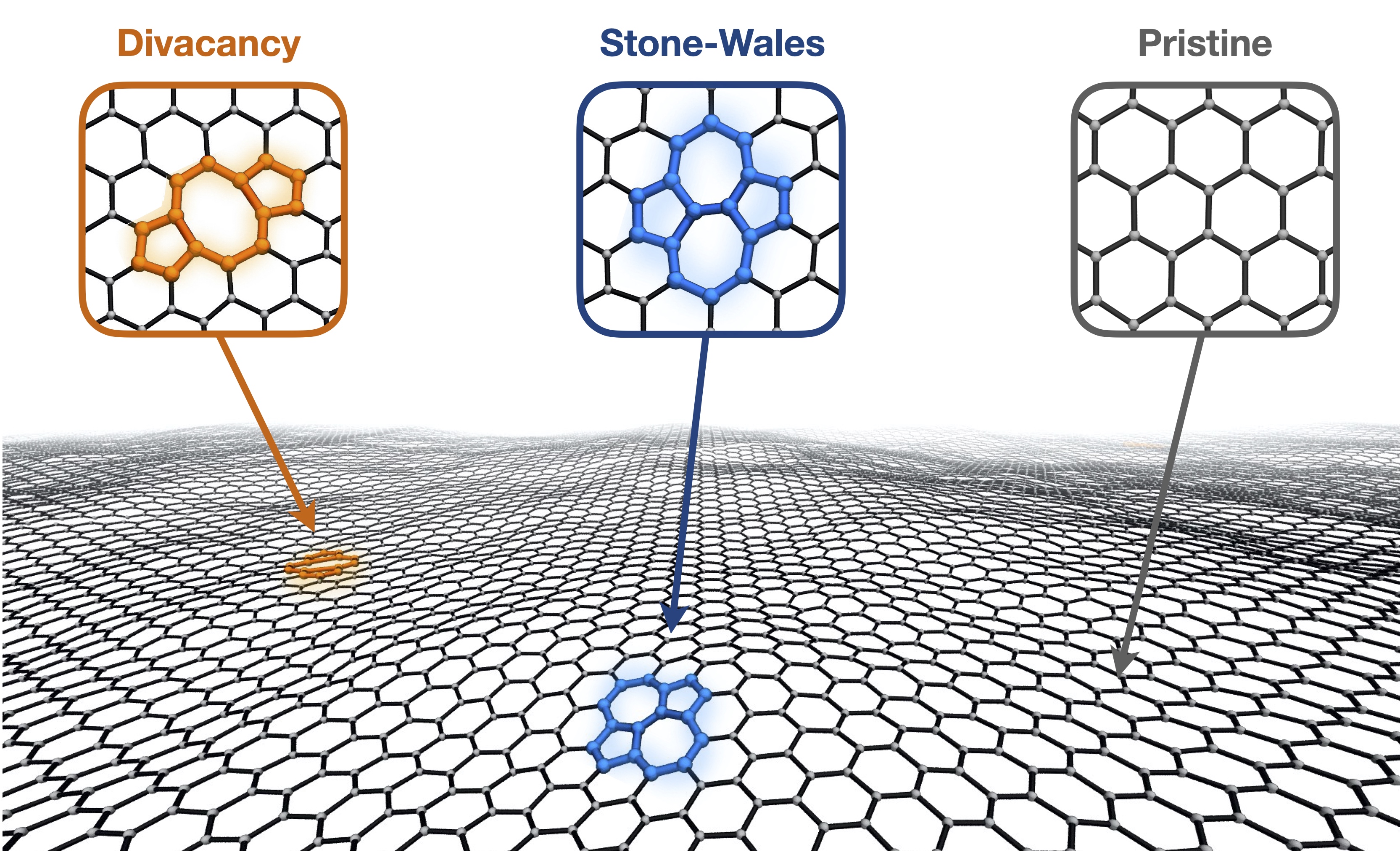}
\caption{{\label{fig:overview}}
Schematic illustration of graphene and the different defect types investigated in this letter and expected morphology in the limit of low defect concentrations. The atoms forming the defect are highlighted in orange and blue for divacancy and Stone-Wales defects, respectively.}
\end{figure}

In this work, we investigate graphene's structural response to two common point defects \cite{Banhart2011, Robertson2013,Skowron2015}, namely divacancies and Stone-Wales \cite{Stone1986} defects both visualised in figure \ref{fig:overview}.
Stone-Wales defects represent the simplest example of a topological defect where a C-C bond rotation transforms four hexagons into two pentagons and heptagons.
Conversely, divacancies are formed by the loss of two adjacent carbon atoms.  
Graphene's $sp^2$~network remains intact, however, by undergoing a reconstruction and the formation of one 8-membered and two 5-membered rings \cite{Kim2011,Robertson2012}. 
It is worth noting, however, that additional bond rotations can further transform the divacancy into more complex geometries comprising a rich variety of geometrical shapes including pentagons, hexagons, and heptagons \cite{Kotakoski2011} (see also supporting information).\\

We employ molecular dynamics (MD) simulations to explore the defect-induced alteration of graphene's structure. 
As the spatial extension of graphene's ripples \cite{Fasolino2007,Thiemann2020} and long-range nature of the defect-defect interactions \cite{Warner2012,Lehtinen2013,Kotakoski2014,Samsonidze2002} make simulation studies with \textit{ab initio} methods prohibitively expensive, classical force fields such as the long-range carbon order potential (LCBOP) \cite{Los2003}, the reactive empirical bond order (REBO II) potential \cite{Stuart2000a}, or the environment dependent interatomic potential (EDIP) \cite{Marks2001}, represent a computationally efficient alternative.
In this work, however, we use the recently developed machine learning-based Gaussian approximation potential for carbon (GAP-20) \cite{Rowe2020b} which has been carefully validated against experimental measurements and quantum mechanical calculations.
Being trained on a database of \textit{ab initio} structures comprising configurations ranging from pristine graphene to amorphous carbon, GAP-20 reliably describes the phonon spectrum of graphene as well as the defects' energetic and structural characteristics.
By using GAP-20, our work represents the first systematic investigation of the interplay between defects and graphene's morphology based on large scale simulations approaching quantum mechanical accuracy.
For a detailed description of GAP-20 including the selection of training data, chosen hyperparameters as well as a comprehensive benchmarking and comparison to other classical force fields the reader is referred to the original reference \cite{Rowe2020b} while an extensive derivation of the theory behind the GAP framework can be found elsewhere \cite{Bartok2010a,Bartok2015}..
In addition, in the supporting information we show that the GAP-20 accurately reproduces its \textit{ab initio} reference for highly defective and corrugated graphene.\\ 

For both defect types, we perform simulations at room temperature for a varying concentrations ranging from $\approx 0.03 \%$ for an isolated defect to $3 \%$ representing a highly defective system where the distance between defects reduces to $\approx 1~\mathrm{nm}$.

As the equilibrium concentration of the defects studied is negligible at room temperature due to their high formation energies \cite{Banhart2011,Skowron2015}, the high concentrations investigated here can be in principle achieved by intentionally introducing defects through a variety of experimental methods described above.
While divacancies are easily created, the controlled preparation of graphene samples comprising a large number of Stone-Wales defects remains experimentally challenging due to their short lifetime under electron beam irradiation \cite{Meyer2008,Kotakoski2011a,Kotakoski2011}.
From a theoretical point of view, however, Stone-Wales defects, being the fundamental example of topological disorder, represent a suitable reference to gain valuable insight into the type-dependent impact of defects on graphene's corrugation.
Further, both Stone-Wales defects and divacancies do not, in contrast to adatoms \cite{Gerber2010}, migrate at room temperature preventing the coalescence or annihilation of defects and, thus, make the comparison between the two types straightforward.\\

We find that the defective systems indeed exhibit a highly crumpled surface, the dimensions of which significantly exceed the expected out-of-plane deviations in pristine graphene, particularly at defect concentrations $> 0.5 \%$. 
On a quantitative level, however, the impact of Stone-Wales and divacancy defects differs considerably with the latter leading to roughly two times higher corrugation at the same defect concentration.
In order to understand this striking result, we examine the local environments of the defects introduced.
Our analysis reveals that the magnitude of the distinct morphological change is an inherent feature of the defect which is determined by its geometry and the extent of defect-defect coupling.\\

The MD simulations were performed in LAMMPS \cite{Plimpton1997} at a temperature of $300$~K and zero stress on free-standing graphene sheets comprising between 6984 and 7200 carbon atoms equivalent to the system size used in our previous work \cite{Thiemann2020}.
The precise number of atoms depends on the number of defects.
In analogy to previous work \cite{Mortazavi2013}, we define the defect concentration as the ratio of removed (2 per divacancy) or rotated (2 per Stone-Wales defect) atoms to the total number of atoms in a pristine graphene sheet.
The defects were randomly distributed on the graphene sheet while satisfying a minimum distance criterion of 10~$\mathrm{\AA}$ between defect centers.
Each simulation was started from a perfectly flat surface and any reconstruction of vacancy defects happened naturally in the simulation and was not enforced by the initial configuration.
To account for the impact of varying orientation and distance between defects, three simulations with different initial defect distributions were conducted for each type and concentration.
All systems were equilibrated for 20~ps before statistics were collected for 150~ps for the defective systems.
As pristine graphene serves as a reference system in our study, it is simulated for 1~ns to ensure a small statistical uncertainty and prevent error propagation when properties of defective systems are related to the pristine case.
In the supporting information, we show that the properties investigated in this work can be obtained with these simulation times showing an error below $10~\%$ with respect to tenfold longer trajectories.
The entire post-processing analysis was done in Python using features from the ASE \cite{HjorthLarsen2017}, MDAnalysis \cite{MichaudAgrawal2011,Gowers2016}, and OVITO \cite{Stukowski2010} software packages.\\


\begin{figure*}[t!]
\centering
\includegraphics[width=0.9\linewidth]{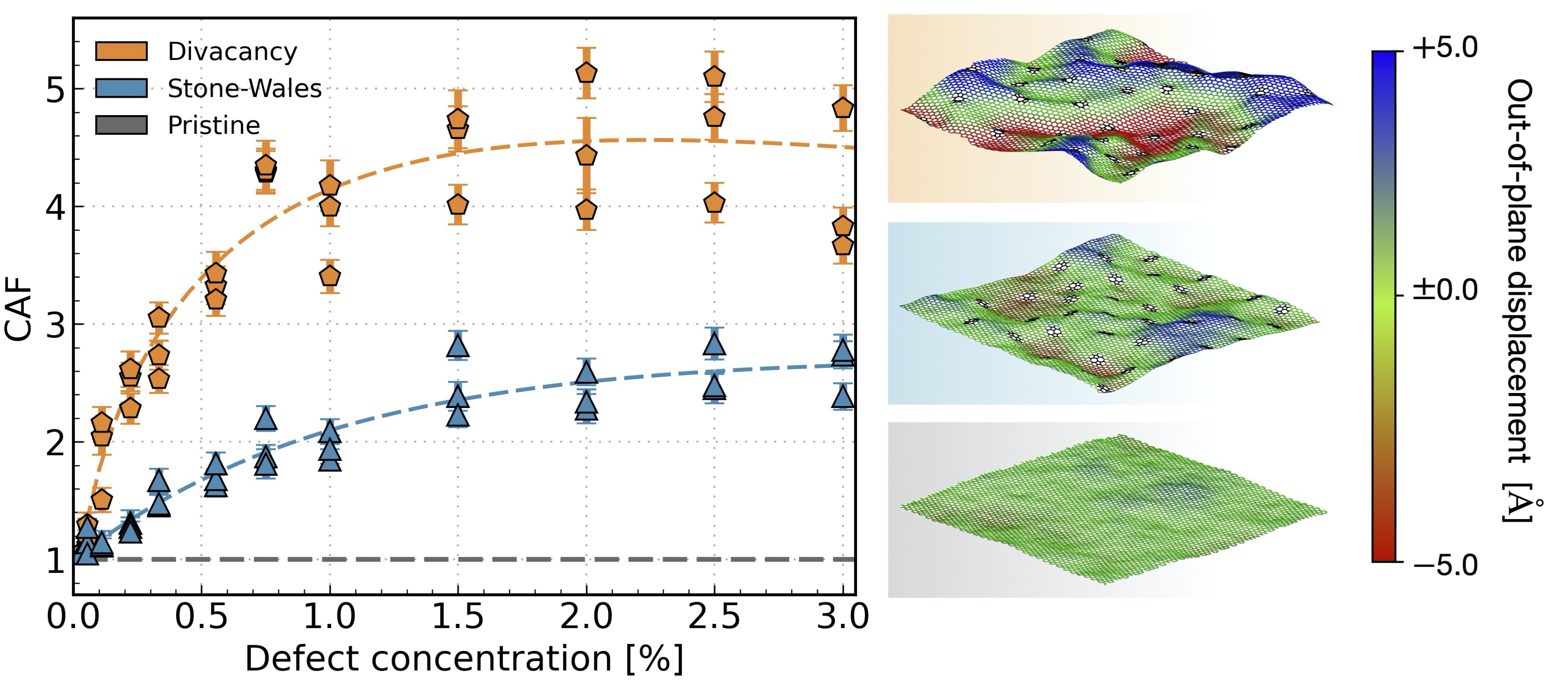}
\caption{{\label{fig:height_distribution}} Morphological transformation of graphene in the presence of different types of defects.
The left panel depicts the dependence of the CAF on defect type and concentration.
The CAF is a ratio of the corrugation of each defective system to pristine graphene and the values reported correspond to the simulation averages over all atoms and frames.
The error bars represent the related statistical error obtained via block averaging.
The dashed lines are intended as guides to the eye.
On the right hand side, we show snapshots from the trajectories at $1 \%$ concentration for the two types of defect and for pristine graphene.
The atoms are coloured according to their out-of-plane position relative to the center of mass of the graphene sheet and the defective atoms are highlighted in black.}
\end{figure*}

We start our investigation by analysing the structure of a graphene sheet in the absence and presence of defects. 
Here, we use the standard deviation of the atomic heights distribution sampled over all atoms and frames as a representative measure to quantify the morphology of the different systems.
Rather than reporting absolute values, however, we express the height fluctuations of any system relative to those observed for pristine graphene and denote this ratio as the corrugation amplification factor (CAF, see also supporting information).
While the atomic out-of-plane displacements intrinsically scale with the system's dimensions \cite{Nelson1987,LeDoussal1992,Nelson2004}, the CAF represents an intuitive and invariant property to assess whether certain defects have an enhancing or diminishing impact on graphene's corrugation.\\

In figure \ref{fig:height_distribution}, we show the CAF as a function of the defect concentration for the two distinct types of structural defects.
The CAF values differ significantly across the individual defect types and concentrations, ranging from approximately 1 (no morphological alteration with respect to pristine graphene) to 5 (substantial topological change).
While the different simulations at equal concentrations of the same type yield remarkably similar results, it is noteworthy that scattering beyond the range of statistical errors of the measured CAFs is inevitable due to different initial defect distributions, particularly at high concentrations. 
The exact orientation and arrangement of defects can fundamentally affect graphene's properties \cite{Wang2013,He2014} and also have been shown to determine which buckling modes are induced \cite{Lehtinen2013}. \\

Graphene comprising Stone-Wales defects or divacancies shows CAF values well above 1 even at relatively low concentrations of $\approx 0.2 \%$, demonstrating their capability to increase the amplitude of the atomic height fluctuations.
For both defect types, we initially observe a strong increase of the CAF before reaching a plateau at high concentrations $ > 1.5 \%$.
At $3 \%$, the highest concentration investigated, graphene's morphology differs fundamentally from the pristine condition by possessing an almost three-times and five-times more corrugated surface for Stone-Wales defects and divacancies, respectively.
This strong dependence on the defect type and concentration is also reflected by the discrepancy of CAF values observed for differently reconstructed divacancies (see supporting information) indicating a high sensitivity of graphene's morphology with respect to small perturbations such as individual bond rotations.
To visualise the dimension of the extreme structural transformations, snapshots from the simulations at a concentration of $1 \%$ are shown for both defect types and compared to pristine graphene on the right panel of figure \ref{fig:height_distribution}.
The defective graphene sheets show significant out-of-plane buckling differing fundamentally from the pristine system. 
Interestingly, the magnitude of this defect-induced corrugation is very similar to that of pristine graphene under compressive strain showing a CAF of $\approx 6$ at a strain rate of $-2 \%$ (see supporting information and references \cite{Rowe2018, Thiemann2020}).
\\

\begin{figure*}[t!]
\centering
\includegraphics[width=0.9\linewidth]{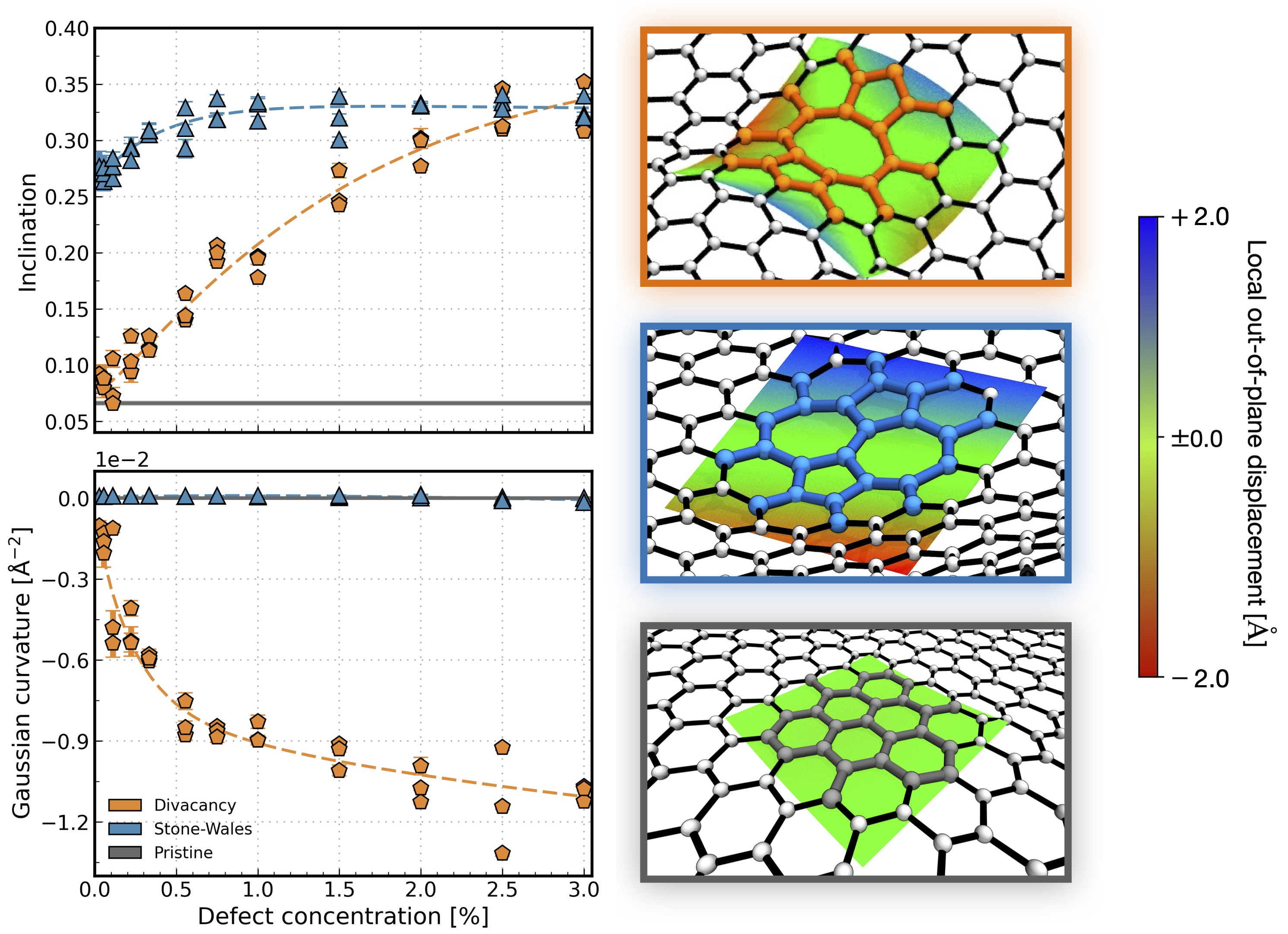}
\caption{{\label{fig:local_environments}} Analysis of the geometrical properties of the defects' local environments.
The left half of the figures shows the results for the local inclination (top) and Gaussian curvature (bottom) for all defect types as a function of concentration.
For each property, the reported values correspond to averages over all defects and frames and the error bars represent the related statistical error obtained via block averaging.
The solid lines correspond to the reference value obtained for pristine graphene and the dashed lines are intended as guides to the eye.
On the right side, we show close-ups of simulation snapshots depicting the different systems: divacancies (orange), Stone-Wales defects (blue), and pristine graphene (grey).
In each snapshot the atoms of the local environment of a defect are highlighted by thicker bonds and the colour of the respective defect type. 
The coloured surfaces correspond to the fitted function $f_h(x,y)$ for the respective defect which are coloured according to their local height relative to the center of mass of the defective atoms. 
}
\end{figure*}

The distinct trends observed in graphene's structural behaviour call for a detailed analysis of the atomic structure in the defects' vicinity to further understand the origin and mechanisms of the morphological transformations. 
To this end, we compute the average inclination and Gaussian curvature of the defects' local environments as shown in the left panel of figure \ref{fig:local_environments}.
Here, we describe these geometrical quantities based on an analytical height function $f_h(x,y)$ which is individually fitted to the atomic positions defining the local environment of each defect in the system.
Having ensured an accurate description of the atomic landscape, first and second order spatial derivatives of $f_h(x,y)$ are computed along previously defined characteristic directions resembling the defect type's unique symmetry.
The inclination and Gaussian curvature are then computed as the norm of the gradient and the determinant of the Hessian at the defect center, respectively.\\
%

To provide a baseline and get a grasp of the expected magnitudes, we evaluate both parameters on pristine graphene where three adjacent hexagons are employed as a defect proxy , as illustrated in the lower right panel of figure \ref{fig:local_environments}.
As anticipated, local environments exclusively comprising hexagons show zero Gaussian curvature and the presence of wrinkles and ripples in pristine graphene can be attributed entirely to local tilting which we quantify here to be $\approx 0.066$.
Note that this non-zero inclination by no means contradicts the general requirement imposed by periodic boundary conditions of a flat surface on average where negative and positive atomic out-of-plane displacements balance each other out.
Rather, it originates from strictly positive values of the geometric measure being based on the norm of the local height gradients and its magnitude is related to the overall surface roughness and corrugation.\\

While Stone-Wales defects show zero local Gaussian curvature, they exhibit a roughly four to five times larger tilt than unperturbed hexagons in pristine graphene.
The high value of $\approx 0.270$ for the isolated defect highlights that the strong inclination is an intrinsic feature of the Stone-Wales defect.
It was shown by \textit{ab initio} calculations \cite{Ma2009} that it is energetically most favourable for the atoms forming the pentagons to move in opposite directions imposing a sine-like wave centered at the defect core. 
We observe the same trend for concentrations $< 0.25 ~\%$ where the total inclination can be almost exclusively attributed to the height gradient along the pentagon axis (see supporting information) as visualised in the center right panel of figure \ref{fig:local_environments}.
A high defect concentration has a positive but rather small effect on the local inclination showing a maximum increase of $< 20 \%$ reaching a plateau for concentrations $>1.5~\%$. \\

With an inclination and Gaussian curvature ranging from $\approx 0.086$ to $\approx 0.33$ and from $\approx -0.001$ to $\approx -0.012$, respectively, divacancies show a concentration dependence for both geometrical parameters.
Based on the definition of the Gaussian curvature, negative values correspond to bending in opposing directions along orthogonal axes corresponding to a pringle \cite{Liu2010} shape (hyperbolic paraboloid) as illustrated in figure \ref{fig:local_environments}.
This curved shape is characteristic for divacancies and agrees well with the minimum energy configurations found with a classical potential \cite{Leyssale2014} and density functional theory \cite{Kotakoski2014}.
In contrast to Stone-Wales defects, however, we find a significant enhancement of both geometrical parameters by increasing the concentration leading to a roughly four and tenfold higher inclination and curvature, respectively, indicating strong defect-defect coupling. \\

Our analysis of the local environments reveals the origin behind the distinct morphological alterations for both defect types.
The defects act as \textit{corrugation seeds} and impose a distinct shift in the height distributions of the adjacent atoms to comply with the induced tilt and curvature, respectively.
For dilute and very low concentrations, this phenomenon is mainly of local nature showing a decaying effect with increasing distance from the defect center resulting in a global morphology similar to that of pristine graphene.
In the abundance of defects, conversely, the perturbations induced by individual \textit{seeds} will superimpose and interfere with each other preventing the overall atomic height distribution from approaching the pristine limit.
This effect can be enhanced if the defects strongly interact with each other as is shown in the case of divacancies.\\

Finally, as this is the first simulation study scanning the morphological impact of different point defects at varying concentrations and room temperature, a direct comparison to previous work is difficult.
While a complete benchmarking of other potentials is beyond the scope of this work, we analysed trajectories obtained from MD simulations performed at defect concentration of $1~\%$ based on the commonly employed LCBOP and REBO II force fields.
Overall, the CAFs obtained with these potentials agree qualitatively with our findings.
However, we note that both force fields overestimate the energy barrier associated with the reconstruction of the divacancy, requiring the simulation to be initialised with reconstructed defects to obtain a corrugated morphology (see supporting information).
Thus, simply using these potentials without prior knowledge of the influence defects have on the corrugation is unlikely to have revealed the insights obtained here.\\


In conclusion, we have reported a machine learning-based molecular dynamics study investigating the impact of Stone-Wales and divacancy defects at varying concentrations and room temperature on the morphological behaviour of graphene.
In accordance with theory and chemical intuition, we find that defects can significantly enhance the system's corrugation even at low concentrations with the magnitude of the influence depending considerably on the nature of the defect. 
To further understand the mechanisms behind these trends, we compared the defects' local structure revealing an intrinsic tilt and curvature for Stone-Wales and divacancy defects, respectively, whereby the latter is strongly enhanced by defect-defect coupling.
Our results underline the strong interplay between defects and graphene's corrugation indicating that the surprising mechanical properties \cite{Lopez-Polin2015,Lopez-Polin2017} of defective graphene might be related to the substantially more wrinkled surface.
Looking forward, in this work we exclusively focused on the structural alteration of free-standing graphene in the presence of defects of the same type.
However, it will be relevant to investigate the coupling between different types of defects as well as the impact of substrate-induced strain on graphene's corrugation in future work.
Likewise, it will be interesting to explore the impact of defects on the dynamics of the graphene ripples which have been shown to couple and dominate the motion of a water droplet on strained graphene \cite{Ma2016}.
By linking the global morphology in graphene to the fundamental nature of each defect type, our work represents a starting point to answering these questions and paves the way for precise nanoengineering of graphene in various applications ranging from the removal of organic pollutants\cite{Wang2014,Wang2016} to lithium ion batteries\cite{Odkhuu2014}.

\section*{Acknowledgements}

We thank Christoph Schran, Matthias Kiesel, Stephen J Cox, and Venkat Kapil for fruitful discussions throughout the course of this work.
We are grateful to the UK Materials and Molecular Modelling Hub for computational resources, which is partially funded by EPSRC (EP/P020194/1 and EP/T022213/1).
This work used the ARCHER UK National Supercomputing Service (\textcolor{blue}{http://www.archer.ac.uk}) through
our membership of the UK’s HEC Materials Chemistry Consortium, which is funded by EPSRC (EP/L000202, EP/ R029431).
We are also grateful for the computational resources granted by the UCL Grace High Performance Computing Facility (Grace@UCL), and associated support services. 
Calculations were also performed on the Cambridge Service for Data Driven Discovery (CSD3) operated by the University of Cambridge Research Computing Service (\textcolor{blue}{www.csd3.cam.ac.uk}), provided by Dell EMC and Intel using Tier-2 funding from the Engineering and Physical Sciences Research Council (capital grant EP/P020259/1), and DiRAC funding from the Science and Technology Facilities Council (\textcolor{blue}{www.dirac.ac.uk}).
AZ acknowledges financial support from the Leverhulme Trust, grant number RPG-2020-038.

\section*{References}

\bibliography{paper.bib}

\providecommand{\latin}[1]{#1}
\makeatletter
\providecommand{\doi}
  {\begingroup\let\do\@makeother\dospecials
  \catcode`\{=1 \catcode`\}=2 \doi@aux}
\providecommand{\doi@aux}[1]{\endgroup\texttt{#1}}
\makeatother
\providecommand*\mcitethebibliography{\thebibliography}
\csname @ifundefined\endcsname{endmcitethebibliography}
  {\let\endmcitethebibliography\endthebibliography}{}
\begin{mcitethebibliography}{104}
\providecommand*\natexlab[1]{#1}
\providecommand*\mciteSetBstSublistMode[1]{}
\providecommand*\mciteSetBstMaxWidthForm[2]{}
\providecommand*\mciteBstWouldAddEndPuncttrue
  {\def\EndOfBibitem{\unskip.}}
\providecommand*\mciteBstWouldAddEndPunctfalse
  {\let\EndOfBibitem\relax}
\providecommand*\mciteSetBstMidEndSepPunct[3]{}
\providecommand*\mciteSetBstSublistLabelBeginEnd[3]{}
\providecommand*\EndOfBibitem{}
\mciteSetBstSublistMode{f}
\mciteSetBstMaxWidthForm{subitem}{(\alph{mcitesubitemcount})}
\mciteSetBstSublistLabelBeginEnd
  {\mcitemaxwidthsubitemform\space}
  {\relax}
  {\relax}

\bibitem[Fasolino \latin{et~al.}(2007)Fasolino, Los, and
  Katsnelson]{Fasolino2007}
Fasolino,~A.; Los,~J.~H.; Katsnelson,~M.~I. {Intrinsic ripples in graphene}.
  \emph{Nature Materials} \textbf{2007}, \emph{6}, 858--861\relax
\mciteBstWouldAddEndPuncttrue
\mciteSetBstMidEndSepPunct{\mcitedefaultmidpunct}
{\mcitedefaultendpunct}{\mcitedefaultseppunct}\relax
\EndOfBibitem
\bibitem[Qin \latin{et~al.}(2016)Qin, Sun, Liu, and Liu]{Qin2016}
Qin,~H.; Sun,~Y.; Liu,~J.~Z.; Liu,~Y. {Mechanical properties of wrinkled
  graphene generated by topological defects}. \emph{Carbon} \textbf{2016},
  \emph{108}, 204--214\relax
\mciteBstWouldAddEndPuncttrue
\mciteSetBstMidEndSepPunct{\mcitedefaultmidpunct}
{\mcitedefaultendpunct}{\mcitedefaultseppunct}\relax
\EndOfBibitem
\bibitem[Guinea \latin{et~al.}(2008)Guinea, Horovitz, and {Le
  Doussal}]{Guinea2008}
Guinea,~F.; Horovitz,~B.; {Le Doussal},~P. {Gauge field induced by ripples in
  graphene}. \emph{Physical Review B - Condensed Matter and Materials Physics}
  \textbf{2008}, \emph{77}, 205421\relax
\mciteBstWouldAddEndPuncttrue
\mciteSetBstMidEndSepPunct{\mcitedefaultmidpunct}
{\mcitedefaultendpunct}{\mcitedefaultseppunct}\relax
\EndOfBibitem
\bibitem[Guinea \latin{et~al.}(2008)Guinea, Katsnelson, and
  Vozmediano]{Guinea2008a}
Guinea,~F.; Katsnelson,~M.~I.; Vozmediano,~M.~A. {Midgap states and charge
  inhomogeneities in corrugated graphene}. \emph{Physical Review B - Condensed
  Matter and Materials Physics} \textbf{2008}, \emph{77}, 075422\relax
\mciteBstWouldAddEndPuncttrue
\mciteSetBstMidEndSepPunct{\mcitedefaultmidpunct}
{\mcitedefaultendpunct}{\mcitedefaultseppunct}\relax
\EndOfBibitem
\bibitem[{V{\'{a}}zquez De Parga} \latin{et~al.}(2008){V{\'{a}}zquez De Parga},
  Calleja, Borca, Passeggi, Hinarejos, Guinea, and Miranda]{VazquezDeParga2008}
{V{\'{a}}zquez De Parga},~A.~L.; Calleja,~F.; Borca,~B.; Passeggi,~M.~C.;
  Hinarejos,~J.~J.; Guinea,~F.; Miranda,~R. {Periodically rippled graphene:
  Growth and spatially resolved electronic structure}. \emph{Physical Review
  Letters} \textbf{2008}, \emph{100}, 056807\relax
\mciteBstWouldAddEndPuncttrue
\mciteSetBstMidEndSepPunct{\mcitedefaultmidpunct}
{\mcitedefaultendpunct}{\mcitedefaultseppunct}\relax
\EndOfBibitem
\bibitem[Boukhvalov and Katsnelson(2009)Boukhvalov, and
  Katsnelson]{Boukhvalov2009}
Boukhvalov,~D.~W.; Katsnelson,~M.~I. {Enhancement of chemical activity in
  corrugated graphene}. \emph{Journal of Physical Chemistry C} \textbf{2009},
  \emph{113}, 14176--14178\relax
\mciteBstWouldAddEndPuncttrue
\mciteSetBstMidEndSepPunct{\mcitedefaultmidpunct}
{\mcitedefaultendpunct}{\mcitedefaultseppunct}\relax
\EndOfBibitem
\bibitem[Qiu \latin{et~al.}(2011)Qiu, Zhang, Yang, Wang, Simon, and
  Li]{Qiu2011}
Qiu,~L.; Zhang,~X.; Yang,~W.; Wang,~Y.; Simon,~G.~P.; Li,~D. {Controllable
  corrugation of chemically converted graphene sheets in water and potential
  application for nanofiltration}. \emph{Chemical Communications}
  \textbf{2011}, \emph{47}, 5810--5812\relax
\mciteBstWouldAddEndPuncttrue
\mciteSetBstMidEndSepPunct{\mcitedefaultmidpunct}
{\mcitedefaultendpunct}{\mcitedefaultseppunct}\relax
\EndOfBibitem
\bibitem[Wei \latin{et~al.}(2014)Wei, Lv, and Xu]{Wei2014}
Wei,~N.; Lv,~C.; Xu,~Z. {Wetting of graphene oxide: A molecular dynamics
  study}. \emph{Langmuir} \textbf{2014}, \emph{30}, 3572--3578\relax
\mciteBstWouldAddEndPuncttrue
\mciteSetBstMidEndSepPunct{\mcitedefaultmidpunct}
{\mcitedefaultendpunct}{\mcitedefaultseppunct}\relax
\EndOfBibitem
\bibitem[Dahanayaka \latin{et~al.}(2017)Dahanayaka, Liu, Hu, Chen, Law, and
  Zhou]{Dahanayaka2017}
Dahanayaka,~M.; Liu,~B.; Hu,~Z.; Chen,~Z.; Law,~A. W.~K.; Zhou,~K. {Corrugated
  graphene layers for sea water desalination using capacitive deionization}.
  \emph{Physical Chemistry Chemical Physics} \textbf{2017}, \emph{19},
  8552--8562\relax
\mciteBstWouldAddEndPuncttrue
\mciteSetBstMidEndSepPunct{\mcitedefaultmidpunct}
{\mcitedefaultendpunct}{\mcitedefaultseppunct}\relax
\EndOfBibitem
\bibitem[Marbach \latin{et~al.}(2018)Marbach, Dean, and Bocquet]{Marbach2018}
Marbach,~S.; Dean,~D.~S.; Bocquet,~L. {Transport and dispersion across wiggling
  nanopores}. \emph{Nature Physics} \textbf{2018}, \emph{14}, 1108--1113\relax
\mciteBstWouldAddEndPuncttrue
\mciteSetBstMidEndSepPunct{\mcitedefaultmidpunct}
{\mcitedefaultendpunct}{\mcitedefaultseppunct}\relax
\EndOfBibitem
\bibitem[Lehtinen \latin{et~al.}(2003)Lehtinen, Foster, Ayuela, Krasheninnikov,
  Nordlund, and Nieminen]{Lehtinen2003}
Lehtinen,~P.~O.; Foster,~A.~S.; Ayuela,~A.; Krasheninnikov,~A.; Nordlund,~K.;
  Nieminen,~R.~M. {Magnetic Properties and Diffusion of Adatoms on a Graphene
  Sheet}. \emph{Physical Review Letters} \textbf{2003}, \emph{91}, 017202\relax
\mciteBstWouldAddEndPuncttrue
\mciteSetBstMidEndSepPunct{\mcitedefaultmidpunct}
{\mcitedefaultendpunct}{\mcitedefaultseppunct}\relax
\EndOfBibitem
\bibitem[Sammalkorpi \latin{et~al.}(2004)Sammalkorpi, Krasheninnikov, Kuronen,
  Nordlund, and Kaski]{Sammalkorpi2004}
Sammalkorpi,~M.; Krasheninnikov,~A.; Kuronen,~A.; Nordlund,~K.; Kaski,~K.
  {Mechanical properties of carbon nanotubes with vacancies and related
  defects}. \emph{Physical Review B - Condensed Matter and Materials Physics}
  \textbf{2004}, \emph{70}, 245416\relax
\mciteBstWouldAddEndPuncttrue
\mciteSetBstMidEndSepPunct{\mcitedefaultmidpunct}
{\mcitedefaultendpunct}{\mcitedefaultseppunct}\relax
\EndOfBibitem
\bibitem[{Costa Filho} \latin{et~al.}(2007){Costa Filho}, Farias, and
  Peeters]{CostaFilho2007}
{Costa Filho},~R.~N.; Farias,~G.~A.; Peeters,~F.~M. {Graphene ribbons with a
  line of impurities: Opening of a gap}. \emph{Physical Review B - Condensed
  Matter and Materials Physics} \textbf{2007}, \emph{76}, 193409\relax
\mciteBstWouldAddEndPuncttrue
\mciteSetBstMidEndSepPunct{\mcitedefaultmidpunct}
{\mcitedefaultendpunct}{\mcitedefaultseppunct}\relax
\EndOfBibitem
\bibitem[Kopylov \latin{et~al.}(2011)Kopylov, Cheianov, Altshuler, and
  Fal'Ko]{Kopylov2011}
Kopylov,~S.~V.; Cheianov,~V.; Altshuler,~B.~L.; Fal'Ko,~V.~I. {Transport
  anomaly at the ordering transition for adatoms on graphene}. \emph{Physical
  Review B - Condensed Matter and Materials Physics} \textbf{2011}, \emph{83},
  201401(R)\relax
\mciteBstWouldAddEndPuncttrue
\mciteSetBstMidEndSepPunct{\mcitedefaultmidpunct}
{\mcitedefaultendpunct}{\mcitedefaultseppunct}\relax
\EndOfBibitem
\bibitem[Wang \latin{et~al.}(2012)Wang, Yan, Ma, Hu, and Chen]{Wang2012}
Wang,~M.~C.; Yan,~C.; Ma,~L.; Hu,~N.; Chen,~M.~W. {Effect of defects on
  fracture strength of graphene sheets}. \emph{Computational Materials Science}
  \textbf{2012}, \emph{54}, 236--239\relax
\mciteBstWouldAddEndPuncttrue
\mciteSetBstMidEndSepPunct{\mcitedefaultmidpunct}
{\mcitedefaultendpunct}{\mcitedefaultseppunct}\relax
\EndOfBibitem
\bibitem[Ansari \latin{et~al.}(2012)Ansari, Ajori, and Motevalli]{Ansari2012}
Ansari,~R.; Ajori,~S.; Motevalli,~B. {Mechanical properties of defective
  single-layered graphene sheets via molecular dynamics simulation}.
  \emph{Superlattices and Microstructures} \textbf{2012}, \emph{51},
  274--289\relax
\mciteBstWouldAddEndPuncttrue
\mciteSetBstMidEndSepPunct{\mcitedefaultmidpunct}
{\mcitedefaultendpunct}{\mcitedefaultseppunct}\relax
\EndOfBibitem
\bibitem[Mortazavi and Ahzi(2013)Mortazavi, and Ahzi]{Mortazavi2013}
Mortazavi,~B.; Ahzi,~S. {Thermal conductivity and tensile response of defective
  graphene: A molecular dynamics study}. \emph{Carbon} \textbf{2013},
  \emph{63}, 460--470\relax
\mciteBstWouldAddEndPuncttrue
\mciteSetBstMidEndSepPunct{\mcitedefaultmidpunct}
{\mcitedefaultendpunct}{\mcitedefaultseppunct}\relax
\EndOfBibitem
\bibitem[He \latin{et~al.}(2014)He, Guo, Lei, Sha, and Liu]{He2014}
He,~L.; Guo,~S.; Lei,~J.; Sha,~Z.; Liu,~Z. {The effect of Stone-Thrower-Wales
  defects on mechanical properties of graphene sheets - A molecular dynamics
  study}. \emph{Carbon} \textbf{2014}, \emph{75}, 124--132\relax
\mciteBstWouldAddEndPuncttrue
\mciteSetBstMidEndSepPunct{\mcitedefaultmidpunct}
{\mcitedefaultendpunct}{\mcitedefaultseppunct}\relax
\EndOfBibitem
\bibitem[Zandiatashbar \latin{et~al.}(2014)Zandiatashbar, Lee, An, Lee, Mathew,
  Terrones, Hayashi, Picu, Hone, and Koratkar]{Zandiatashbar2014}
Zandiatashbar,~A.; Lee,~G.~H.; An,~S.~J.; Lee,~S.; Mathew,~N.; Terrones,~M.;
  Hayashi,~T.; Picu,~C.~R.; Hone,~J.; Koratkar,~N. {Effect of defects on the
  intrinsic strength and stiffness of graphene}. \emph{Nature Communications}
  \textbf{2014}, \emph{5}, 3186\relax
\mciteBstWouldAddEndPuncttrue
\mciteSetBstMidEndSepPunct{\mcitedefaultmidpunct}
{\mcitedefaultendpunct}{\mcitedefaultseppunct}\relax
\EndOfBibitem
\bibitem[Wei \latin{et~al.}(2012)Wei, Wu, Yin, Shi, Yang, and
  Dresselhaus]{Wei2012}
Wei,~Y.; Wu,~J.; Yin,~H.; Shi,~X.; Yang,~R.; Dresselhaus,~M. {The nature of
  strength enhancement and weakening by pentagong-heptagon defects in
  graphene}. \emph{Nature Materials} \textbf{2012}, \emph{11}, 759--763\relax
\mciteBstWouldAddEndPuncttrue
\mciteSetBstMidEndSepPunct{\mcitedefaultmidpunct}
{\mcitedefaultendpunct}{\mcitedefaultseppunct}\relax
\EndOfBibitem
\bibitem[Boukhvalov and Katsnelson(2008)Boukhvalov, and
  Katsnelson]{Boukhvalov2008}
Boukhvalov,~D.~W.; Katsnelson,~M.~I. {Chemical functionalization of graphene
  with defects}. \emph{Nano Letters} \textbf{2008}, \emph{8}, 4374--4379\relax
\mciteBstWouldAddEndPuncttrue
\mciteSetBstMidEndSepPunct{\mcitedefaultmidpunct}
{\mcitedefaultendpunct}{\mcitedefaultseppunct}\relax
\EndOfBibitem
\bibitem[Banhart \latin{et~al.}(2011)Banhart, Kotakoski, and
  Krasheninnikov]{Banhart2011}
Banhart,~F.; Kotakoski,~J.; Krasheninnikov,~A.~V. {Structural defects in
  graphene}. \emph{ACS Nano} \textbf{2011}, \emph{5}, 26--41\relax
\mciteBstWouldAddEndPuncttrue
\mciteSetBstMidEndSepPunct{\mcitedefaultmidpunct}
{\mcitedefaultendpunct}{\mcitedefaultseppunct}\relax
\EndOfBibitem
\bibitem[Nair \latin{et~al.}(2012)Nair, Sepioni, Tsai, Lehtinen, Keinonen,
  Krasheninnikov, Thomson, Geim, and Grigorieva]{Nair2012}
Nair,~R.~R.; Sepioni,~M.; Tsai,~I.~L.; Lehtinen,~O.; Keinonen,~J.;
  Krasheninnikov,~A.~V.; Thomson,~T.; Geim,~A.~K.; Grigorieva,~I.~V. {Spin-half
  paramagnetism in graphene induced by point defects}. \emph{Nature Physics}
  \textbf{2012}, \emph{8}, 199--202\relax
\mciteBstWouldAddEndPuncttrue
\mciteSetBstMidEndSepPunct{\mcitedefaultmidpunct}
{\mcitedefaultendpunct}{\mcitedefaultseppunct}\relax
\EndOfBibitem
\bibitem[Malekpour \latin{et~al.}(2016)Malekpour, Ramnani, Srinivasan,
  Balasubramanian, Nika, Mulchandani, Lake, and Balandin]{Malekpour2016}
Malekpour,~H.; Ramnani,~P.; Srinivasan,~S.; Balasubramanian,~G.; Nika,~D.~L.;
  Mulchandani,~A.; Lake,~R.~K.; Balandin,~A.~A. {Thermal conductivity of
  graphene with defects induced by electron beam irradiation}. \emph{Nanoscale}
  \textbf{2016}, \emph{8}, 14608--14616\relax
\mciteBstWouldAddEndPuncttrue
\mciteSetBstMidEndSepPunct{\mcitedefaultmidpunct}
{\mcitedefaultendpunct}{\mcitedefaultseppunct}\relax
\EndOfBibitem
\bibitem[Valencia and Caldas(2017)Valencia, and Caldas]{Valencia2017}
Valencia,~A.~M.; Caldas,~M.~J. {Single vacancy defect in graphene: Insights
  into its magnetic properties from theoretical modeling}. \emph{Physical
  Review B} \textbf{2017}, \emph{96}, 125431\relax
\mciteBstWouldAddEndPuncttrue
\mciteSetBstMidEndSepPunct{\mcitedefaultmidpunct}
{\mcitedefaultendpunct}{\mcitedefaultseppunct}\relax
\EndOfBibitem
\bibitem[Griffin \latin{et~al.}(2020)Griffin, Mogg, Hao, Hao, Kalon, Kalon,
  Bacaksiz, Lopez-Polin, Lopez-Polin, Zhou, Guarochico, Cai, Neumann, Winter,
  Mohn, Lee, Lin, Lin, Kaiser, Grigorieva, Suenaga, {\"{O}}zyilmaz, Cheng,
  Cheng, Ren, Turchanin, Peeters, Geim, and Lozada-Hidalgo]{Griffin2020}
Griffin,~E. \latin{et~al.}  {Proton and Li-Ion Permeation through Graphene with
  Eight-Atom-Ring Defects}. \emph{ACS Nano} \textbf{2020}, \emph{14},
  7280--7286\relax
\mciteBstWouldAddEndPuncttrue
\mciteSetBstMidEndSepPunct{\mcitedefaultmidpunct}
{\mcitedefaultendpunct}{\mcitedefaultseppunct}\relax
\EndOfBibitem
\bibitem[Hu \latin{et~al.}(2014)Hu, Lozada-Hidalgo, Wang, Mishchenko, Schedin,
  Nair, Hill, Boukhvalov, Katsnelson, Dryfe, Grigorieva, Wu, and Geim]{Hu2014a}
Hu,~S.; Lozada-Hidalgo,~M.; Wang,~F.~C.; Mishchenko,~A.; Schedin,~F.;
  Nair,~R.~R.; Hill,~E.~W.; Boukhvalov,~D.~W.; Katsnelson,~M.~I.; Dryfe,~R.~A.;
  Grigorieva,~I.~V.; Wu,~H.~A.; Geim,~A.~K. {Proton transport through
  one-atom-thick crystals}. \emph{Nature} \textbf{2014}, \emph{516},
  227--230\relax
\mciteBstWouldAddEndPuncttrue
\mciteSetBstMidEndSepPunct{\mcitedefaultmidpunct}
{\mcitedefaultendpunct}{\mcitedefaultseppunct}\relax
\EndOfBibitem
\bibitem[Sun \latin{et~al.}(2020)Sun, Yang, Kuang, Stebunov, Xiong, Yu, Nair,
  Katsnelson, Yuan, Grigorieva, Lozada-Hidalgo, Wang, and Geim]{Sun2020}
Sun,~P.~Z.; Yang,~Q.; Kuang,~W.~J.; Stebunov,~Y.~V.; Xiong,~W.~Q.; Yu,~J.;
  Nair,~R.~R.; Katsnelson,~M.~I.; Yuan,~S.~J.; Grigorieva,~I.~V.;
  Lozada-Hidalgo,~M.; Wang,~F.~C.; Geim,~A.~K. {Limits on gas impermeability of
  graphene}. \emph{Nature} \textbf{2020}, \emph{579}, 229--232\relax
\mciteBstWouldAddEndPuncttrue
\mciteSetBstMidEndSepPunct{\mcitedefaultmidpunct}
{\mcitedefaultendpunct}{\mcitedefaultseppunct}\relax
\EndOfBibitem
\bibitem[Lusk and Carr(2008)Lusk, and Carr]{Lusk2008}
Lusk,~M.~T.; Carr,~L.~D. {Nanoengineering defect structures on graphene}.
  \emph{Physical Review Letters} \textbf{2008}, \emph{100}, 175503\relax
\mciteBstWouldAddEndPuncttrue
\mciteSetBstMidEndSepPunct{\mcitedefaultmidpunct}
{\mcitedefaultendpunct}{\mcitedefaultseppunct}\relax
\EndOfBibitem
\bibitem[Krasheninnikov and Banhart(2007)Krasheninnikov, and
  Banhart]{Krasheninnikov2007}
Krasheninnikov,~A.~V.; Banhart,~F. {Engineering of nanostructured carbon
  materials with electron or ion beams}. \emph{Nature Materials} \textbf{2007},
  \emph{6}, 723--733\relax
\mciteBstWouldAddEndPuncttrue
\mciteSetBstMidEndSepPunct{\mcitedefaultmidpunct}
{\mcitedefaultendpunct}{\mcitedefaultseppunct}\relax
\EndOfBibitem
\bibitem[Krasheninnikov \latin{et~al.}(2001)Krasheninnikov, Nordlund,
  Sirvi{\"{o}}, and {Salonen, E.Keinonen}]{Krasheninnikov2001}
Krasheninnikov,~A.~V.; Nordlund,~K.; Sirvi{\"{o}},~M.; {Salonen,
  E.Keinonen},~J. {Formation of ion-irradiation-induced atomic-scale defects on
  walls of carbon nanotubes}. \emph{Physical Review B - Condensed Matter and
  Materials Physics} \textbf{2001}, \emph{63}, 245405\relax
\mciteBstWouldAddEndPuncttrue
\mciteSetBstMidEndSepPunct{\mcitedefaultmidpunct}
{\mcitedefaultendpunct}{\mcitedefaultseppunct}\relax
\EndOfBibitem
\bibitem[Kotakoski \latin{et~al.}(2011)Kotakoski, Krasheninnikov, Kaiser, and
  Meyer]{Kotakoski2011}
Kotakoski,~J.; Krasheninnikov,~A.~V.; Kaiser,~U.; Meyer,~J.~C. {From point
  defects in graphene to two-dimensional amorphous carbon}. \emph{Physical
  Review Letters} \textbf{2011}, \emph{106}, 105505\relax
\mciteBstWouldAddEndPuncttrue
\mciteSetBstMidEndSepPunct{\mcitedefaultmidpunct}
{\mcitedefaultendpunct}{\mcitedefaultseppunct}\relax
\EndOfBibitem
\bibitem[Kotakoski \latin{et~al.}(2011)Kotakoski, Meyer, Kurasch,
  Santos-Cottin, Kaiser, and Krasheninnikov]{Kotakoski2011a}
Kotakoski,~J.; Meyer,~J.~C.; Kurasch,~S.; Santos-Cottin,~D.; Kaiser,~U.;
  Krasheninnikov,~A.~V. {Stone-Wales-type transformations in carbon
  nanostructures driven by electron irradiation}. \emph{Physical Review B -
  Condensed Matter and Materials Physics} \textbf{2011}, \emph{83},
  245420\relax
\mciteBstWouldAddEndPuncttrue
\mciteSetBstMidEndSepPunct{\mcitedefaultmidpunct}
{\mcitedefaultendpunct}{\mcitedefaultseppunct}\relax
\EndOfBibitem
\bibitem[Robertson \latin{et~al.}(2012)Robertson, Allen, Wu, He, Olivier,
  Neethling, Kirkland, and Warner]{Robertson2012}
Robertson,~A.~W.; Allen,~C.~S.; Wu,~Y.~A.; He,~K.; Olivier,~J.; Neethling,~J.;
  Kirkland,~A.~I.; Warner,~J.~H. {Spatial control of defect creation in
  graphene at the nanoscale}. \emph{Nature Communications} \textbf{2012},
  \emph{3}, 1144\relax
\mciteBstWouldAddEndPuncttrue
\mciteSetBstMidEndSepPunct{\mcitedefaultmidpunct}
{\mcitedefaultendpunct}{\mcitedefaultseppunct}\relax
\EndOfBibitem
\bibitem[Susi \latin{et~al.}(2017)Susi, Meyer, and Kotakoski]{Susi2017}
Susi,~T.; Meyer,~J.~C.; Kotakoski,~J. {Manipulating low-dimensional materials
  down to the level of single atoms with electron irradiation}.
  \emph{Ultramicroscopy} \textbf{2017}, \emph{180}, 163--172\relax
\mciteBstWouldAddEndPuncttrue
\mciteSetBstMidEndSepPunct{\mcitedefaultmidpunct}
{\mcitedefaultendpunct}{\mcitedefaultseppunct}\relax
\EndOfBibitem
\bibitem[Zhao \latin{et~al.}(2017)Zhao, Kotakoski, Meyer, Sutter, Sutter,
  Krasheninnikov, Kaiser, and Zhou]{Zhao2017}
Zhao,~X.; Kotakoski,~J.; Meyer,~J.~C.; Sutter,~E.; Sutter,~P.;
  Krasheninnikov,~A.~V.; Kaiser,~U.; Zhou,~W. {Engineering and modifying
  two-dimensional materials by electron beams}. \emph{MRS Bulletin}
  \textbf{2017}, \emph{42}, 667--676\relax
\mciteBstWouldAddEndPuncttrue
\mciteSetBstMidEndSepPunct{\mcitedefaultmidpunct}
{\mcitedefaultendpunct}{\mcitedefaultseppunct}\relax
\EndOfBibitem
\bibitem[Tripathi \latin{et~al.}(2018)Tripathi, Mittelberger, Pike, Mangler,
  Meyer, Verstraete, Kotakoski, and Susi]{Tripathi2018}
Tripathi,~M.; Mittelberger,~A.; Pike,~N.~A.; Mangler,~C.; Meyer,~J.~C.;
  Verstraete,~M.~J.; Kotakoski,~J.; Susi,~T. {Electron-Beam Manipulation of
  Silicon Dopants in Graphene}. \emph{Nano Letters} \textbf{2018}, \emph{18},
  5319--5323\relax
\mciteBstWouldAddEndPuncttrue
\mciteSetBstMidEndSepPunct{\mcitedefaultmidpunct}
{\mcitedefaultendpunct}{\mcitedefaultseppunct}\relax
\EndOfBibitem
\bibitem[Trentino \latin{et~al.}(2021)Trentino, Madsen, Mittelberger, Mangler,
  Susi, Mustonen, and Kotakoski]{Trentino2021}
Trentino,~A.; Madsen,~J.; Mittelberger,~A.; Mangler,~C.; Susi,~T.;
  Mustonen,~K.; Kotakoski,~J. {Atomic-Level Structural Engineering of Graphene
  on a Mesoscopic Scale}. \emph{Nano Letters} \textbf{2021}, \emph{21},
  5179--5185\relax
\mciteBstWouldAddEndPuncttrue
\mciteSetBstMidEndSepPunct{\mcitedefaultmidpunct}
{\mcitedefaultendpunct}{\mcitedefaultseppunct}\relax
\EndOfBibitem
\bibitem[Yazyev \latin{et~al.}(2007)Yazyev, Tavernelli, Rothlisberger, and
  Helm]{Yazyev2007}
Yazyev,~O.~V.; Tavernelli,~I.; Rothlisberger,~U.; Helm,~L. {Early stages of
  radiation damage in graphite and carbon nanostructures: A first-principles
  molecular dynamics study}. \emph{Physical Review B - Condensed Matter and
  Materials Physics} \textbf{2007}, \emph{75}, 115418\relax
\mciteBstWouldAddEndPuncttrue
\mciteSetBstMidEndSepPunct{\mcitedefaultmidpunct}
{\mcitedefaultendpunct}{\mcitedefaultseppunct}\relax
\EndOfBibitem
\bibitem[Yoon \latin{et~al.}(2016)Yoon, Rahnamoun, Swett, Iberi, Cullen,
  Vlassiouk, Belianinov, Jesse, Sang, Ovchinnikova, Rondinone, Unocic, and {Van
  Duin}]{Yoon2016}
Yoon,~K.; Rahnamoun,~A.; Swett,~J.~L.; Iberi,~V.; Cullen,~D.~A.;
  Vlassiouk,~I.~V.; Belianinov,~A.; Jesse,~S.; Sang,~X.; Ovchinnikova,~O.~S.;
  Rondinone,~A.~J.; Unocic,~R.~R.; {Van Duin},~A.~C. {Atomistic-Scale
  Simulations of Defect Formation in Graphene under Noble Gas Ion Irradiation}.
  \emph{ACS Nano} \textbf{2016}, \emph{10}, 8376--8384\relax
\mciteBstWouldAddEndPuncttrue
\mciteSetBstMidEndSepPunct{\mcitedefaultmidpunct}
{\mcitedefaultendpunct}{\mcitedefaultseppunct}\relax
\EndOfBibitem
\bibitem[Shi \latin{et~al.}(2019)Shi, Peng, Bai, Gao, and Jovanovic]{Shi2019}
Shi,~T.; Peng,~Q.; Bai,~Z.; Gao,~F.; Jovanovic,~I. {Proton irradiation of
  graphene: Insights from atomistic modeling}. \emph{Nanoscale} \textbf{2019},
  \emph{11}, 20754--20765\relax
\mciteBstWouldAddEndPuncttrue
\mciteSetBstMidEndSepPunct{\mcitedefaultmidpunct}
{\mcitedefaultendpunct}{\mcitedefaultseppunct}\relax
\EndOfBibitem
\bibitem[Su and Xue(2021)Su, and Xue]{Su2021}
Su,~S.; Xue,~J. {Facile Fabrication of Subnanopores in Graphene under Ion
  Irradiation: Molecular Dynamics Simulations}. \emph{ACS Applied Materials and
  Interfaces} \textbf{2021}, \emph{13}, 12366--12374\relax
\mciteBstWouldAddEndPuncttrue
\mciteSetBstMidEndSepPunct{\mcitedefaultmidpunct}
{\mcitedefaultendpunct}{\mcitedefaultseppunct}\relax
\EndOfBibitem
\bibitem[Zhu \latin{et~al.}(2011)Zhu, Murali, Stoller, Ganesh, Cai, Ferreira,
  Pirkle, Wallace, Cychosz, Thommes, Su, Stach, and Ruoff]{Zhu2011}
Zhu,~Y.; Murali,~S.; Stoller,~M.~D.; Ganesh,~K.~J.; Cai,~W.; Ferreira,~P.~J.;
  Pirkle,~A.; Wallace,~R.~M.; Cychosz,~K.~A.; Thommes,~M.; Su,~D.;
  Stach,~E.~A.; Ruoff,~R.~S. {Carbon-Based Supercapacitors}. \emph{Science}
  \textbf{2011}, \emph{332}, 1537--1542\relax
\mciteBstWouldAddEndPuncttrue
\mciteSetBstMidEndSepPunct{\mcitedefaultmidpunct}
{\mcitedefaultendpunct}{\mcitedefaultseppunct}\relax
\EndOfBibitem
\bibitem[Wang \latin{et~al.}(2016)Wang, Chen, and Xing]{Wang2016}
Wang,~J.; Chen,~B.; Xing,~B. {Wrinkles and Folds of Activated Graphene
  Nanosheets as Fast and Efficient Adsorptive Sites for Hydrophobic Organic
  Contaminants}. \emph{Environmental Science and Technology} \textbf{2016},
  \emph{50}, 3798--3808\relax
\mciteBstWouldAddEndPuncttrue
\mciteSetBstMidEndSepPunct{\mcitedefaultmidpunct}
{\mcitedefaultendpunct}{\mcitedefaultseppunct}\relax
\EndOfBibitem
\bibitem[Bao \latin{et~al.}(2009)Bao, Miao, Chen, Zhang, Jang, Dames, and
  Lau]{Bao2009}
Bao,~W.; Miao,~F.; Chen,~Z.; Zhang,~H.; Jang,~W.; Dames,~C.; Lau,~C.~N.
  {Controlled ripple texturing of suspended graphene and ultrathin graphite
  membranes}. \emph{Nature Nanotechnology} \textbf{2009}, \emph{4},
  562--566\relax
\mciteBstWouldAddEndPuncttrue
\mciteSetBstMidEndSepPunct{\mcitedefaultmidpunct}
{\mcitedefaultendpunct}{\mcitedefaultseppunct}\relax
\EndOfBibitem
\bibitem[Toh \latin{et~al.}(2020)Toh, Zhang, Lin, Mayorov, Wang, Orofeo, Ferry,
  Andersen, Kakenov, Guo, Abidi, Sims, Suenaga, Pantelides, and
  {\"{O}}zyilmaz]{Toh2020}
Toh,~C.~T.; Zhang,~H.; Lin,~J.; Mayorov,~A.~S.; Wang,~Y.~P.; Orofeo,~C.~M.;
  Ferry,~D.~B.; Andersen,~H.; Kakenov,~N.; Guo,~Z.; Abidi,~I.~H.; Sims,~H.;
  Suenaga,~K.; Pantelides,~S.~T.; {\"{O}}zyilmaz,~B. {Synthesis and properties
  of free-standing monolayer amorphous carbon}. \emph{Nature} \textbf{2020},
  \emph{577}, 199--203\relax
\mciteBstWouldAddEndPuncttrue
\mciteSetBstMidEndSepPunct{\mcitedefaultmidpunct}
{\mcitedefaultendpunct}{\mcitedefaultseppunct}\relax
\EndOfBibitem
\bibitem[Zhao \latin{et~al.}(2019)Zhao, Xu, Ma, Liu, Zhou, Liu, Feng, Zhu,
  Kang, Sun, Cheng, and Ren]{Zhao2019}
Zhao,~T.; Xu,~C.; Ma,~W.; Liu,~Z.; Zhou,~T.; Liu,~Z.; Feng,~S.; Zhu,~M.;
  Kang,~N.; Sun,~D.~M.; Cheng,~H.~M.; Ren,~W. {Ultrafast growth of
  nanocrystalline graphene films by quenching and grain-size-dependent strength
  and bandgap opening}. \emph{Nature Communications} \textbf{2019}, \emph{10},
  4854\relax
\mciteBstWouldAddEndPuncttrue
\mciteSetBstMidEndSepPunct{\mcitedefaultmidpunct}
{\mcitedefaultendpunct}{\mcitedefaultseppunct}\relax
\EndOfBibitem
\bibitem[Tozzini and Pellegrini(2011)Tozzini, and Pellegrini]{Tozzini2011}
Tozzini,~V.; Pellegrini,~V. {Reversible hydrogen storage by controlled buckling
  of graphene layers}. \emph{Journal of Physical Chemistry C} \textbf{2011},
  \emph{115}, 25523--25528\relax
\mciteBstWouldAddEndPuncttrue
\mciteSetBstMidEndSepPunct{\mcitedefaultmidpunct}
{\mcitedefaultendpunct}{\mcitedefaultseppunct}\relax
\EndOfBibitem
\bibitem[Guo \latin{et~al.}(2012)Guo, Morris, Ihm, Contescu, Gallego, Duscher,
  Pennycook, and Chisholm]{Guo2012}
Guo,~J.; Morris,~J.~R.; Ihm,~Y.; Contescu,~C.~I.; Gallego,~N.~C.; Duscher,~G.;
  Pennycook,~S.~J.; Chisholm,~M.~F. {Topological defects: Origin of nanopores
  and enhanced adsorption performance in nanoporous carbon}. \emph{Small}
  \textbf{2012}, \emph{8}, 3283--3288\relax
\mciteBstWouldAddEndPuncttrue
\mciteSetBstMidEndSepPunct{\mcitedefaultmidpunct}
{\mcitedefaultendpunct}{\mcitedefaultseppunct}\relax
\EndOfBibitem
\bibitem[Ito \latin{et~al.}(2016)Ito, Shen, Hojo, Itagaki, Fujita, Chen, Aida,
  Tang, Adschiri, and Chen]{Ito2016}
Ito,~Y.; Shen,~Y.; Hojo,~D.; Itagaki,~Y.; Fujita,~T.; Chen,~L.; Aida,~T.;
  Tang,~Z.; Adschiri,~T.; Chen,~M. {Correlation between Chemical Dopants and
  Topological Defects in Catalytically Active Nanoporous Graphene}.
  \emph{Advanced Materials} \textbf{2016}, \emph{28}, 10644--10651\relax
\mciteBstWouldAddEndPuncttrue
\mciteSetBstMidEndSepPunct{\mcitedefaultmidpunct}
{\mcitedefaultendpunct}{\mcitedefaultseppunct}\relax
\EndOfBibitem
\bibitem[Wang \latin{et~al.}(2014)Wang, Chen, and Chen]{Wang2014}
Wang,~J.; Chen,~Z.; Chen,~B. {Adsorption of polycyclic aromatic hydrocarbons by
  graphene and graphene oxide nanosheets}. \emph{Environmental Science and
  Technology} \textbf{2014}, \emph{48}, 4817--4825\relax
\mciteBstWouldAddEndPuncttrue
\mciteSetBstMidEndSepPunct{\mcitedefaultmidpunct}
{\mcitedefaultendpunct}{\mcitedefaultseppunct}\relax
\EndOfBibitem
\bibitem[Glukhova and Slepchenkov(2012)Glukhova, and Slepchenkov]{Glukhova2012}
Glukhova,~O.; Slepchenkov,~M. {Influence of the curvature of deformed graphene
  nanoribbons on their electronic and adsorptive properties: Theoretical
  investigation based on the analysis of the local stress field for an atomic
  grid}. \emph{Nanoscale} \textbf{2012}, \emph{4}, 3335--3344\relax
\mciteBstWouldAddEndPuncttrue
\mciteSetBstMidEndSepPunct{\mcitedefaultmidpunct}
{\mcitedefaultendpunct}{\mcitedefaultseppunct}\relax
\EndOfBibitem
\bibitem[Chen and Chen(2015)Chen, and Chen]{Chen2015}
Chen,~X.; Chen,~B. {Macroscopic and spectroscopic investigations of the
  adsorption of nitroaromatic compounds on graphene oxide, reduced graphene
  oxide, and graphene nanosheets}. \emph{Environmental Science and Technology}
  \textbf{2015}, \emph{49}, 6181--6189\relax
\mciteBstWouldAddEndPuncttrue
\mciteSetBstMidEndSepPunct{\mcitedefaultmidpunct}
{\mcitedefaultendpunct}{\mcitedefaultseppunct}\relax
\EndOfBibitem
\bibitem[Radich and Kamat(2013)Radich, and Kamat]{Radich2013}
Radich,~J.~G.; Kamat,~P.~V. {Making graphene Holey. Gold-nanoparticle-mediated
  hydroxyl radical attack on reduced graphene oxide}. \emph{ACS Nano}
  \textbf{2013}, \emph{7}, 5546--5557\relax
\mciteBstWouldAddEndPuncttrue
\mciteSetBstMidEndSepPunct{\mcitedefaultmidpunct}
{\mcitedefaultendpunct}{\mcitedefaultseppunct}\relax
\EndOfBibitem
\bibitem[Terrones and Mackay(1992)Terrones, and Mackay]{Terrones1992}
Terrones,~H.; Mackay,~A.~L. {The geometry of hypothetical curved graphite
  structures}. \emph{Carbon} \textbf{1992}, \emph{30}, 1251--1260\relax
\mciteBstWouldAddEndPuncttrue
\mciteSetBstMidEndSepPunct{\mcitedefaultmidpunct}
{\mcitedefaultendpunct}{\mcitedefaultseppunct}\relax
\EndOfBibitem
\bibitem[Irvine \latin{et~al.}(2010)Irvine, Vitelli, and Chaikin]{Irvine2010}
Irvine,~W.~T.; Vitelli,~V.; Chaikin,~P.~M. {Pleats in crystals on curved
  surfaces}. \emph{Nature} \textbf{2010}, \emph{468}, 947--951\relax
\mciteBstWouldAddEndPuncttrue
\mciteSetBstMidEndSepPunct{\mcitedefaultmidpunct}
{\mcitedefaultendpunct}{\mcitedefaultseppunct}\relax
\EndOfBibitem
\bibitem[Liu and Yakobson(2010)Liu, and Yakobson]{Liu2010}
Liu,~Y.; Yakobson,~B.~I. {Cones, pringles, and grain boundary landscapes in
  graphene topology}. \emph{Nano Letters} \textbf{2010}, \emph{10},
  2178--2183\relax
\mciteBstWouldAddEndPuncttrue
\mciteSetBstMidEndSepPunct{\mcitedefaultmidpunct}
{\mcitedefaultendpunct}{\mcitedefaultseppunct}\relax
\EndOfBibitem
\bibitem[Kawasumi \latin{et~al.}(2013)Kawasumi, Zhang, Segawa, Scott, and
  Itami]{Kawasumi2013}
Kawasumi,~K.; Zhang,~Q.; Segawa,~Y.; Scott,~L.~T.; Itami,~K. {A grossly warped
  nanographene and the consequences of multiple odd-membered-ring defects}.
  \emph{Nature Chemistry} \textbf{2013}, \emph{5}, 739--744\relax
\mciteBstWouldAddEndPuncttrue
\mciteSetBstMidEndSepPunct{\mcitedefaultmidpunct}
{\mcitedefaultendpunct}{\mcitedefaultseppunct}\relax
\EndOfBibitem
\bibitem[Kusumaatmaja and Wales(2013)Kusumaatmaja, and Wales]{Kusumaatmaja2013}
Kusumaatmaja,~H.; Wales,~D.~J. {Defect motifs for constant mean curvature
  surfaces}. \emph{Physical Review Letters} \textbf{2013}, \emph{110},
  165502\relax
\mciteBstWouldAddEndPuncttrue
\mciteSetBstMidEndSepPunct{\mcitedefaultmidpunct}
{\mcitedefaultendpunct}{\mcitedefaultseppunct}\relax
\EndOfBibitem
\bibitem[Brojan \latin{et~al.}(2015)Brojan, Terwagne, Lagrange, and
  Reis]{Brojan2015}
Brojan,~M.; Terwagne,~D.; Lagrange,~R.; Reis,~P.~M. {Wrinkling crystallography
  on spherical surfaces}. \emph{Proceedings of the National Academy of Sciences
  of the United States of America} \textbf{2015}, \emph{112}, 14--19\relax
\mciteBstWouldAddEndPuncttrue
\mciteSetBstMidEndSepPunct{\mcitedefaultmidpunct}
{\mcitedefaultendpunct}{\mcitedefaultseppunct}\relax
\EndOfBibitem
\bibitem[Warner \latin{et~al.}(2013)Warner, Fan, Robertson, He, Yoon, and
  Lee]{Warner2013}
Warner,~J.~H.; Fan,~Y.; Robertson,~A.~W.; He,~K.; Yoon,~E.; Lee,~G.~D.
  {Rippling graphene at the nanoscale through dislocation addition}. \emph{Nano
  Letters} \textbf{2013}, \emph{13}, 4937--4944\relax
\mciteBstWouldAddEndPuncttrue
\mciteSetBstMidEndSepPunct{\mcitedefaultmidpunct}
{\mcitedefaultendpunct}{\mcitedefaultseppunct}\relax
\EndOfBibitem
\bibitem[Lehtinen \latin{et~al.}(2013)Lehtinen, Kurasch, Krasheninnikov, and
  Kaiser]{Lehtinen2013}
Lehtinen,~O.; Kurasch,~S.; Krasheninnikov,~A.~V.; Kaiser,~U. {Atomic scale
  study of the life cycle of a dislocation in graphene from birth to
  annihilation}. \emph{Nature Communications} \textbf{2013}, \emph{4},
  2098\relax
\mciteBstWouldAddEndPuncttrue
\mciteSetBstMidEndSepPunct{\mcitedefaultmidpunct}
{\mcitedefaultendpunct}{\mcitedefaultseppunct}\relax
\EndOfBibitem
\bibitem[Jain \latin{et~al.}(2015)Jain, Barkema, Mousseau, Fang, and {Van
  Huis}]{Jain2015}
Jain,~S.~K.; Barkema,~G.~T.; Mousseau,~N.; Fang,~C.~M.; {Van Huis},~M.~A.
  {Strong Long-Range Relaxations of Structural Defects in Graphene Simulated
  Using a New Semiempirical Potential}. \emph{Journal of Physical Chemistry C}
  \textbf{2015}, \emph{119}, 9646--9655\relax
\mciteBstWouldAddEndPuncttrue
\mciteSetBstMidEndSepPunct{\mcitedefaultmidpunct}
{\mcitedefaultendpunct}{\mcitedefaultseppunct}\relax
\EndOfBibitem
\bibitem[Schniepp \latin{et~al.}(2008)Schniepp, Kudin, Li, Prud'Homme, Car,
  Saville, and Aksay]{Schniepp2008}
Schniepp,~H.~C.; Kudin,~K.~N.; Li,~J.~L.; Prud'Homme,~R.~K.; Car,~R.;
  Saville,~D.~A.; Aksay,~I.~A. {Bending properties of single functionalized
  graphene sheets probed by atomic force microscopy}. \emph{ACS Nano}
  \textbf{2008}, \emph{2}, 2577--2584\relax
\mciteBstWouldAddEndPuncttrue
\mciteSetBstMidEndSepPunct{\mcitedefaultmidpunct}
{\mcitedefaultendpunct}{\mcitedefaultseppunct}\relax
\EndOfBibitem
\bibitem[Kotakoski \latin{et~al.}(2014)Kotakoski, Eder, and
  Meyer]{Kotakoski2014}
Kotakoski,~J.; Eder,~F.~R.; Meyer,~J.~C. {Atomic structure and energetics of
  large vacancies in graphene}. \emph{Physical Review B - Condensed Matter and
  Materials Physics} \textbf{2014}, \emph{89}, 201406(R)\relax
\mciteBstWouldAddEndPuncttrue
\mciteSetBstMidEndSepPunct{\mcitedefaultmidpunct}
{\mcitedefaultendpunct}{\mcitedefaultseppunct}\relax
\EndOfBibitem
\bibitem[Samsonidze \latin{et~al.}(2002)Samsonidze, Samsonidze, and
  Yakobson]{Samsonidze2002}
Samsonidze,~G.~G.; Samsonidze,~G.~G.; Yakobson,~B.~I. {Energetics of
  Stone-Wales defects in deformations of monoatomic hexagonal layers}.
  \emph{Computational Materials Science} \textbf{2002}, \emph{23}, 62--72\relax
\mciteBstWouldAddEndPuncttrue
\mciteSetBstMidEndSepPunct{\mcitedefaultmidpunct}
{\mcitedefaultendpunct}{\mcitedefaultseppunct}\relax
\EndOfBibitem
\bibitem[Pang \latin{et~al.}(2019)Pang, Deng, Liu, Peng, and Wei]{Pang2019}
Pang,~Z.; Deng,~B.; Liu,~Z.; Peng,~H.; Wei,~Y. {Defects guided wrinkling in
  graphene on copper substrate}. \emph{Carbon} \textbf{2019}, \emph{143},
  736--742\relax
\mciteBstWouldAddEndPuncttrue
\mciteSetBstMidEndSepPunct{\mcitedefaultmidpunct}
{\mcitedefaultendpunct}{\mcitedefaultseppunct}\relax
\EndOfBibitem
\bibitem[Lehtinen \latin{et~al.}(2015)Lehtinen, Vats, Algara-Siller, Knyrim,
  and Kaiser]{Lehtinen2015}
Lehtinen,~O.; Vats,~N.; Algara-Siller,~G.; Knyrim,~P.; Kaiser,~U. {Implantation
  and atomic-scale investigation of self-interstitials in graphene}. \emph{Nano
  Letters} \textbf{2015}, \emph{15}, 235--241\relax
\mciteBstWouldAddEndPuncttrue
\mciteSetBstMidEndSepPunct{\mcitedefaultmidpunct}
{\mcitedefaultendpunct}{\mcitedefaultseppunct}\relax
\EndOfBibitem
\bibitem[Yazyev and Louie(2010)Yazyev, and Louie]{Yazyev2010}
Yazyev,~O.~V.; Louie,~S.~G. {Topological defects in graphene: Dislocations and
  grain boundaries}. \emph{Physical Review B - Condensed Matter and Materials
  Physics} \textbf{2010}, \emph{81}, 195420\relax
\mciteBstWouldAddEndPuncttrue
\mciteSetBstMidEndSepPunct{\mcitedefaultmidpunct}
{\mcitedefaultendpunct}{\mcitedefaultseppunct}\relax
\EndOfBibitem
\bibitem[Liu \latin{et~al.}(2011)Liu, Gajewski, Pao, and Chang]{Liu2011}
Liu,~T.~H.; Gajewski,~G.; Pao,~C.~W.; Chang,~C.~C. {Structure, energy, and
  structural transformations of graphene grain boundaries from atomistic
  simulations}. \emph{Carbon} \textbf{2011}, \emph{49}, 2306--2317\relax
\mciteBstWouldAddEndPuncttrue
\mciteSetBstMidEndSepPunct{\mcitedefaultmidpunct}
{\mcitedefaultendpunct}{\mcitedefaultseppunct}\relax
\EndOfBibitem
\bibitem[Hofer \latin{et~al.}(2018)Hofer, Kramberger, Monazam, Mangler,
  Mittelberger, Argentero, Kotakoski, and Meyer]{Hofer2018}
Hofer,~C.; Kramberger,~C.; Monazam,~M. R.~A.; Mangler,~C.; Mittelberger,~A.;
  Argentero,~G.; Kotakoski,~J.; Meyer,~J.~C. {Revealing the 3d structure of
  graphene defects}. \emph{2D Materials} \textbf{2018}, \emph{5}, 045029\relax
\mciteBstWouldAddEndPuncttrue
\mciteSetBstMidEndSepPunct{\mcitedefaultmidpunct}
{\mcitedefaultendpunct}{\mcitedefaultseppunct}\relax
\EndOfBibitem
\bibitem[Warner \latin{et~al.}(2012)Warner, Margine, Mukai, Robertson,
  Giustino, and Kirkland]{Warner2012}
Warner,~J.~H.; Margine,~E.~R.; Mukai,~M.; Robertson,~A.~W.; Giustino,~F.;
  Kirkland,~A.~I. {Dislocation-Driven Deformations in Graphene}. \emph{Science}
  \textbf{2012}, \emph{337}, 209--212\relax
\mciteBstWouldAddEndPuncttrue
\mciteSetBstMidEndSepPunct{\mcitedefaultmidpunct}
{\mcitedefaultendpunct}{\mcitedefaultseppunct}\relax
\EndOfBibitem
\bibitem[Wang \latin{et~al.}(2013)Wang, Lan, Liu, and Tan]{Wang2013}
Wang,~C.~G.; Lan,~L.; Liu,~Y.~P.; Tan,~H.~F. {Defect-guided wrinkling in
  graphene}. \emph{Computational Materials Science} \textbf{2013}, \emph{77},
  250--253\relax
\mciteBstWouldAddEndPuncttrue
\mciteSetBstMidEndSepPunct{\mcitedefaultmidpunct}
{\mcitedefaultendpunct}{\mcitedefaultseppunct}\relax
\EndOfBibitem
\bibitem[Zhang \latin{et~al.}(2014)Zhang, Li, and Gao]{Zhang2014}
Zhang,~T.; Li,~X.; Gao,~H. {Defects controlled wrinkling and topological design
  in graphene}. \emph{Journal of the Mechanics and Physics of Solids}
  \textbf{2014}, \emph{67}, 2--13\relax
\mciteBstWouldAddEndPuncttrue
\mciteSetBstMidEndSepPunct{\mcitedefaultmidpunct}
{\mcitedefaultendpunct}{\mcitedefaultseppunct}\relax
\EndOfBibitem
\bibitem[Zhang \latin{et~al.}(2014)Zhang, Li, and Gao]{Zhang2014a}
Zhang,~T.; Li,~X.; Gao,~H. {Designing graphene structures with controlled
  distributions of topological defects: A case study of toughness enhancement
  in graphene ruga}. \emph{Extreme Mechanics Letters} \textbf{2014}, \emph{1},
  3--8\relax
\mciteBstWouldAddEndPuncttrue
\mciteSetBstMidEndSepPunct{\mcitedefaultmidpunct}
{\mcitedefaultendpunct}{\mcitedefaultseppunct}\relax
\EndOfBibitem
\bibitem[L{\'{o}}pez-Pol{\'{i}}n \latin{et~al.}(2014)L{\'{o}}pez-Pol{\'{i}}n,
  G{\'{o}}mez-Navarro, Parente, Guinea, Katsnelson, P{\'{e}}rez-Murano, and
  G{\'{o}}mez-Herrero]{Lopez-Polin2015}
L{\'{o}}pez-Pol{\'{i}}n,~G.; G{\'{o}}mez-Navarro,~C.; Parente,~V.; Guinea,~F.;
  Katsnelson,~M.~I.; P{\'{e}}rez-Murano,~F.; G{\'{o}}mez-Herrero,~J.
  {Increasing the elastic modulus of graphene by controlled defect creation}.
  \emph{Nature Physics} \textbf{2014}, \emph{11}, 26--31\relax
\mciteBstWouldAddEndPuncttrue
\mciteSetBstMidEndSepPunct{\mcitedefaultmidpunct}
{\mcitedefaultendpunct}{\mcitedefaultseppunct}\relax
\EndOfBibitem
\bibitem[L{\'{o}}pez-Pol{\'{i}}n \latin{et~al.}(2017)L{\'{o}}pez-Pol{\'{i}}n,
  Ortega, Vilhena, Alda, Gomez-Herrero, Serena, Gomez-Navarro, and
  P{\'{e}}rez]{Lopez-Polin2017}
L{\'{o}}pez-Pol{\'{i}}n,~G.; Ortega,~M.; Vilhena,~J.~G.; Alda,~I.;
  Gomez-Herrero,~J.; Serena,~P.~A.; Gomez-Navarro,~C.; P{\'{e}}rez,~R.
  {Tailoring the thermal expansion of graphene via controlled defect creation}.
  \emph{Carbon} \textbf{2017}, \emph{116}, 670--677\relax
\mciteBstWouldAddEndPuncttrue
\mciteSetBstMidEndSepPunct{\mcitedefaultmidpunct}
{\mcitedefaultendpunct}{\mcitedefaultseppunct}\relax
\EndOfBibitem
\bibitem[Robertson and Warner(2013)Robertson, and Warner]{Robertson2013}
Robertson,~A.~W.; Warner,~J.~H. {Atomic resolution imaging of graphene by
  transmission electron microscopy}. \emph{Nanoscale} \textbf{2013}, \emph{5},
  4079--4093\relax
\mciteBstWouldAddEndPuncttrue
\mciteSetBstMidEndSepPunct{\mcitedefaultmidpunct}
{\mcitedefaultendpunct}{\mcitedefaultseppunct}\relax
\EndOfBibitem
\bibitem[Skowron \latin{et~al.}(2015)Skowron, Lebedeva, Popov, and
  Bichoutskaia]{Skowron2015}
Skowron,~S.~T.; Lebedeva,~I.~V.; Popov,~A.~M.; Bichoutskaia,~E. {Energetics of
  atomic scale structure changes in graphene}. \emph{Chemical Society Reviews}
  \textbf{2015}, \emph{44}, 3143--3176\relax
\mciteBstWouldAddEndPuncttrue
\mciteSetBstMidEndSepPunct{\mcitedefaultmidpunct}
{\mcitedefaultendpunct}{\mcitedefaultseppunct}\relax
\EndOfBibitem
\bibitem[Stone and Wales(1986)Stone, and Wales]{Stone1986}
Stone,~A.~J.; Wales,~D.~J. {Theoretical studies of icosahedral C$_{60}$ and
  some related species}. \emph{Chemical Physics Letters} \textbf{1986},
  \emph{128}, 501--503\relax
\mciteBstWouldAddEndPuncttrue
\mciteSetBstMidEndSepPunct{\mcitedefaultmidpunct}
{\mcitedefaultendpunct}{\mcitedefaultseppunct}\relax
\EndOfBibitem
\bibitem[Kim \latin{et~al.}(2011)Kim, Ihm, Yoon, and Lee]{Kim2011}
Kim,~Y.; Ihm,~J.; Yoon,~E.; Lee,~G.~D. {Dynamics and stability of divacancy
  defects in graphene}. \emph{Physical Review B - Condensed Matter and
  Materials Physics} \textbf{2011}, \emph{84}, 075445\relax
\mciteBstWouldAddEndPuncttrue
\mciteSetBstMidEndSepPunct{\mcitedefaultmidpunct}
{\mcitedefaultendpunct}{\mcitedefaultseppunct}\relax
\EndOfBibitem
\bibitem[Thiemann \latin{et~al.}(2020)Thiemann, Rowe, M{\"{u}}ller, and
  Michaelides]{Thiemann2020}
Thiemann,~F.~L.; Rowe,~P.; M{\"{u}}ller,~E.~A.; Michaelides,~A. {Machine
  Learning Potential for Hexagonal Boron Nitride Applied to Thermally and
  Mechanically Induced Rippling}. \emph{The Journal of Physical Chemistry C}
  \textbf{2020}, \emph{124}, 22278--22290\relax
\mciteBstWouldAddEndPuncttrue
\mciteSetBstMidEndSepPunct{\mcitedefaultmidpunct}
{\mcitedefaultendpunct}{\mcitedefaultseppunct}\relax
\EndOfBibitem
\bibitem[Los and Fasolino(2003)Los, and Fasolino]{Los2003}
Los,~H.; Fasolino,~A. {Intrinsic long-range bond-order potential for carbon:
  Performance in Monte Carlo simulations of graphitization}. \emph{Physical
  Review B - Condensed Matter and Materials Physics} \textbf{2003}, \emph{68},
  024107\relax
\mciteBstWouldAddEndPuncttrue
\mciteSetBstMidEndSepPunct{\mcitedefaultmidpunct}
{\mcitedefaultendpunct}{\mcitedefaultseppunct}\relax
\EndOfBibitem
\bibitem[Stuart \latin{et~al.}(2000)Stuart, Tutein, and Harrison]{Stuart2000a}
Stuart,~S.~J.; Tutein,~A.~B.; Harrison,~J.~A. {A reactive potential for
  hydrocarbons with intermolecular interactions}. \emph{Journal of Chemical
  Physics} \textbf{2000}, \emph{112}, 6472--6486\relax
\mciteBstWouldAddEndPuncttrue
\mciteSetBstMidEndSepPunct{\mcitedefaultmidpunct}
{\mcitedefaultendpunct}{\mcitedefaultseppunct}\relax
\EndOfBibitem
\bibitem[Marks(2001)]{Marks2001}
Marks,~N.~A. {Generalizing the environment-dependent interaction potential for
  carbon}. \emph{Physical Review B - Condensed Matter and Materials Physics}
  \textbf{2001}, \emph{63}, 035401\relax
\mciteBstWouldAddEndPuncttrue
\mciteSetBstMidEndSepPunct{\mcitedefaultmidpunct}
{\mcitedefaultendpunct}{\mcitedefaultseppunct}\relax
\EndOfBibitem
\bibitem[Rowe \latin{et~al.}(2020)Rowe, Deringer, Gasparotto, Cs{\'{a}}nyi, and
  Michaelides]{Rowe2020b}
Rowe,~P.; Deringer,~V.~L.; Gasparotto,~P.; Cs{\'{a}}nyi,~G.; Michaelides,~A.
  {An accurate and transferable machine learning potential for carbon}.
  \emph{Journal of Chemical Physics} \textbf{2020}, \emph{153}, 034702\relax
\mciteBstWouldAddEndPuncttrue
\mciteSetBstMidEndSepPunct{\mcitedefaultmidpunct}
{\mcitedefaultendpunct}{\mcitedefaultseppunct}\relax
\EndOfBibitem
\bibitem[Bart{\'{o}}k \latin{et~al.}(2010)Bart{\'{o}}k, Payne, Kondor, and
  Cs{\'{a}}nyi]{Bartok2010a}
Bart{\'{o}}k,~A.~P.; Payne,~M.~C.; Kondor,~R.; Cs{\'{a}}nyi,~G. {Gaussian
  approximation potentials: The accuracy of quantum mechanics, without the
  electrons}. \emph{Physical Review Letters} \textbf{2010}, \emph{104},
  136403\relax
\mciteBstWouldAddEndPuncttrue
\mciteSetBstMidEndSepPunct{\mcitedefaultmidpunct}
{\mcitedefaultendpunct}{\mcitedefaultseppunct}\relax
\EndOfBibitem
\bibitem[Bart{\'{o}}k and Cs{\'{a}}nyi(2015)Bart{\'{o}}k, and
  Cs{\'{a}}nyi]{Bartok2015}
Bart{\'{o}}k,~A.~P.; Cs{\'{a}}nyi,~G. {Gaussian approximation potentials: A
  brief tutorial introduction}. \emph{International Journal of Quantum
  Chemistry} \textbf{2015}, \emph{115}, 1051--1057\relax
\mciteBstWouldAddEndPuncttrue
\mciteSetBstMidEndSepPunct{\mcitedefaultmidpunct}
{\mcitedefaultendpunct}{\mcitedefaultseppunct}\relax
\EndOfBibitem
\bibitem[Meyer \latin{et~al.}(2008)Meyer, Kisielowski, Erni, Rossell, Crommie,
  and Zettl]{Meyer2008}
Meyer,~J.~C.; Kisielowski,~C.; Erni,~R.; Rossell,~M.~D.; Crommie,~M.~F.;
  Zettl,~A. {Direct imaging of lattice atoms and topological defects in
  graphene membranes}. \emph{Nano Letters} \textbf{2008}, \emph{8},
  3582--3586\relax
\mciteBstWouldAddEndPuncttrue
\mciteSetBstMidEndSepPunct{\mcitedefaultmidpunct}
{\mcitedefaultendpunct}{\mcitedefaultseppunct}\relax
\EndOfBibitem
\bibitem[Gerber \latin{et~al.}(2010)Gerber, Krasheninnikov, Foster, and
  Nieminen]{Gerber2010}
Gerber,~I.~C.; Krasheninnikov,~A.~V.; Foster,~A.~S.; Nieminen,~R.~M. {A
  first-principles study on magnetic coupling between carbon adatoms on
  graphene}. \emph{New Journal of Physics} \textbf{2010}, \emph{12},
  113021\relax
\mciteBstWouldAddEndPuncttrue
\mciteSetBstMidEndSepPunct{\mcitedefaultmidpunct}
{\mcitedefaultendpunct}{\mcitedefaultseppunct}\relax
\EndOfBibitem
\bibitem[Plimpton(1995)]{Plimpton1997}
Plimpton,~S. {Fast Parallel Algorithms for Short-Range Molecular Dynamics}.
  \emph{Journal of Computational Physics} \textbf{1995}, \emph{117},
  1--19\relax
\mciteBstWouldAddEndPuncttrue
\mciteSetBstMidEndSepPunct{\mcitedefaultmidpunct}
{\mcitedefaultendpunct}{\mcitedefaultseppunct}\relax
\EndOfBibitem
\bibitem[{Hjorth Larsen} \latin{et~al.}(2017){Hjorth Larsen}, {J{\O}rgen
  Mortensen}, Blomqvist, Castelli, Christensen, Du{\l}ak, Friis, Groves,
  Hammer, Hargus, Hermes, Jennings, {Bjerre Jensen}, Kermode, Kitchin,
  {Leonhard Kolsbjerg}, Kubal, Kaasbjerg, Lysgaard, {Bergmann Maronsson},
  Maxson, Olsen, Pastewka, Peterson, Rostgaard, Schi{\O}tz, Sch{\"{u}}tt,
  Strange, Thygesen, Vegge, Vilhelmsen, Walter, Zeng, and
  Jacobsen]{HjorthLarsen2017}
{Hjorth Larsen},~A. \latin{et~al.}  {The atomic simulation environment - A
  Python library for working with atoms}. \emph{Journal of Physics Condensed
  Matter} \textbf{2017}, \emph{29}, 273002\relax
\mciteBstWouldAddEndPuncttrue
\mciteSetBstMidEndSepPunct{\mcitedefaultmidpunct}
{\mcitedefaultendpunct}{\mcitedefaultseppunct}\relax
\EndOfBibitem
\bibitem[Michaud-Agrawal \latin{et~al.}(2011)Michaud-Agrawal, Denning, Woolf,
  and Beckstein]{MichaudAgrawal2011}
Michaud-Agrawal,~N.; Denning,~E.~J.; Woolf,~T.~B.; Beckstein,~O. {Software News
  and Updates MDAnalysis: A Toolkit for the Analysis of Molecular Dynamics
  Simulations}. \emph{Journal of computational chemistry} \textbf{2011},
  \emph{32}, 2319--2327\relax
\mciteBstWouldAddEndPuncttrue
\mciteSetBstMidEndSepPunct{\mcitedefaultmidpunct}
{\mcitedefaultendpunct}{\mcitedefaultseppunct}\relax
\EndOfBibitem
\bibitem[Gowers \latin{et~al.}(2016)Gowers, Linke, Barnoud, Reddy, Melo,
  Seyler, Doma{\'{n}}ski, Dotson, Buchoux, Kenney, and Beckstein]{Gowers2016}
Gowers,~R.; Linke,~M.; Barnoud,~J.; Reddy,~T.; Melo,~M.; Seyler,~S.;
  Doma{\'{n}}ski,~J.; Dotson,~D.; Buchoux,~S.; Kenney,~I.; Beckstein,~O.
  {MDAnalysis: A Python Package for the Rapid Analysis of Molecular Dynamics
  Simulations}. \emph{Proceedings of the 15th Python in Science Conference}
  \textbf{2016}, 98--105\relax
\mciteBstWouldAddEndPuncttrue
\mciteSetBstMidEndSepPunct{\mcitedefaultmidpunct}
{\mcitedefaultendpunct}{\mcitedefaultseppunct}\relax
\EndOfBibitem
\bibitem[Stukowski(2010)]{Stukowski2010}
Stukowski,~A. {Visualization and analysis of atomistic simulation data with
  OVITO-the Open Visualization Tool}. \emph{Modelling and Simulation in
  Materials Science and Engineering} \textbf{2010}, \emph{18}, 015012\relax
\mciteBstWouldAddEndPuncttrue
\mciteSetBstMidEndSepPunct{\mcitedefaultmidpunct}
{\mcitedefaultendpunct}{\mcitedefaultseppunct}\relax
\EndOfBibitem
\bibitem[Nelson and Peliti(1987)Nelson, and Peliti]{Nelson1987}
Nelson,~D.~R.; Peliti,~L. {Fluctuations in Membranes With Crystalline and
  Hexatic Order.} \emph{Journal de physique Paris} \textbf{1987}, \emph{48},
  1085--1092\relax
\mciteBstWouldAddEndPuncttrue
\mciteSetBstMidEndSepPunct{\mcitedefaultmidpunct}
{\mcitedefaultendpunct}{\mcitedefaultseppunct}\relax
\EndOfBibitem
\bibitem[{Le Doussal} and Radzihovsky(1992){Le Doussal}, and
  Radzihovsky]{LeDoussal1992}
{Le Doussal},~P.; Radzihovsky,~L. {Self-consistent theory of polymerized
  membranes}. \emph{Physical Review Letters} \textbf{1992}, \emph{69},
  1209--1212\relax
\mciteBstWouldAddEndPuncttrue
\mciteSetBstMidEndSepPunct{\mcitedefaultmidpunct}
{\mcitedefaultendpunct}{\mcitedefaultseppunct}\relax
\EndOfBibitem
\bibitem[Nelson \latin{et~al.}(2004)Nelson, Piran, and Weinberg]{Nelson2004}
Nelson,~D.~R., Piran,~T., Weinberg,~S., Eds. \emph{{Statistical Mechanics of
  Membranes and Surfaces}}, 2nd ed.; World Scientific: Singapore, 2004\relax
\mciteBstWouldAddEndPuncttrue
\mciteSetBstMidEndSepPunct{\mcitedefaultmidpunct}
{\mcitedefaultendpunct}{\mcitedefaultseppunct}\relax
\EndOfBibitem
\bibitem[Rowe \latin{et~al.}(2018)Rowe, Cs{\'{a}}nyi, Alf{\`{e}}, and
  Michaelides]{Rowe2018}
Rowe,~P.; Cs{\'{a}}nyi,~G.; Alf{\`{e}},~D.; Michaelides,~A. {Development of a
  machine learning potential for graphene}. \emph{Physical Review B}
  \textbf{2018}, \emph{97}, 054303\relax
\mciteBstWouldAddEndPuncttrue
\mciteSetBstMidEndSepPunct{\mcitedefaultmidpunct}
{\mcitedefaultendpunct}{\mcitedefaultseppunct}\relax
\EndOfBibitem
\bibitem[Ma \latin{et~al.}(2009)Ma, Alf{\`{e}}, Michaelides, and Wang]{Ma2009}
Ma,~J.; Alf{\`{e}},~D.; Michaelides,~A.; Wang,~E. {Stone-Wales defects in
  graphene and other planar s p2 -bonded materials}. \emph{Physical Review B -
  Condensed Matter and Materials Physics} \textbf{2009}, \emph{80},
  033407\relax
\mciteBstWouldAddEndPuncttrue
\mciteSetBstMidEndSepPunct{\mcitedefaultmidpunct}
{\mcitedefaultendpunct}{\mcitedefaultseppunct}\relax
\EndOfBibitem
\bibitem[Leyssale and Vignoles(2014)Leyssale, and Vignoles]{Leyssale2014}
Leyssale,~J.~M.; Vignoles,~G.~L. {A large-scale molecular dynamics study of the
  divacancy defect in graphene}. \emph{Journal of Physical Chemistry C}
  \textbf{2014}, \emph{118}, 8200--8216\relax
\mciteBstWouldAddEndPuncttrue
\mciteSetBstMidEndSepPunct{\mcitedefaultmidpunct}
{\mcitedefaultendpunct}{\mcitedefaultseppunct}\relax
\EndOfBibitem
\bibitem[Ma \latin{et~al.}(2016)Ma, Tocci, Michaelides, and Aeppli]{Ma2016}
Ma,~M.; Tocci,~G.; Michaelides,~A.; Aeppli,~G. {Fast diffusion of water
  nanodroplets on graphene}. \emph{Nature Materials} \textbf{2016}, \emph{15},
  66--71\relax
\mciteBstWouldAddEndPuncttrue
\mciteSetBstMidEndSepPunct{\mcitedefaultmidpunct}
{\mcitedefaultendpunct}{\mcitedefaultseppunct}\relax
\EndOfBibitem
\bibitem[Odkhuu \latin{et~al.}(2014)Odkhuu, Jung, Lee, Han, Choi, Ruoff, and
  Park]{Odkhuu2014}
Odkhuu,~D.; Jung,~D.~H.; Lee,~H.; Han,~S.~S.; Choi,~S.~H.; Ruoff,~R.~S.;
  Park,~N. {Negatively curved carbon as the anode for lithium ion batteries}.
  \emph{Carbon} \textbf{2014}, \emph{66}, 39--47\relax
\mciteBstWouldAddEndPuncttrue
\mciteSetBstMidEndSepPunct{\mcitedefaultmidpunct}
{\mcitedefaultendpunct}{\mcitedefaultseppunct}\relax
\EndOfBibitem
\end{mcitethebibliography}

\end{document}